\documentclass[doublespacing,reviewcopy,authoryear,12pt]{elsarticle}

\usepackage{amssymb}
\usepackage{amsmath} 
\usepackage{amsfonts} 
\usepackage{amssymb}  
\usepackage{color}
\usepackage{multirow}
\usepackage{graphicx}
\usepackage{topcapt}
\usepackage{multirow}
\usepackage{subfigure}
\usepackage{rotating}
\usepackage{supertabular}
\usepackage{booktabs}
\usepackage[large]{caption}
\usepackage{lineno}
\usepackage{setspace}
\journal{Icarus}

\begin{document}

\begin{frontmatter}


\title{The escape of heavy atoms from the ionosphere of HD209458b. I. A photochemical-dynamical model of the thermosphere.}

\author[rvt]{T. T. Koskinen\corref{cor1}}
\ead{tommi@lpl.arizona.edu}
\cortext[cor1]{Corresponding author. Fax: +1 (520) 621 4933}
\address[rvt]{Lunar and Planetary Laboratory, University of Arizona, 1629 E. University Blvd., Tucson, AZ 85721, USA}

\author[ucl]{M. J. Harris}
\address[ucl]{Department of Physics and Astronomy, University College London, Gower Street, London WC1E 6BT, UK}

\author[rvt]{R. V. Yelle}

\author[france]{P. Lavvas}
\address[france]{Groupe de Spectrom\'etrie Mol\'eculaire et Atmosph\'erique UMR CNRS 6089, Universit\'e Reims Champagne-Ardenne, 51687, France}

\begin{abstract}
The detections of atomic hydrogen, heavy atoms and ions surrounding the extrasolar giant planet (EGP) HD209458b constrain the composition, temperature and density profiles in its upper atmosphere.  Thus the observations provide guidance for models that have so far predicted a range of possible conditions.  We present the first hydrodynamic escape model for the upper atmosphere that includes all of the detected species in order to explain their presence at high altitudes, and to further constrain the temperature and velocity profiles.  This model calculates the stellar heating rates based on recent estimates of photoelectron heating efficiencies, and includes the photochemistry of heavy atoms and ions in addition to hydrogen and helium.  The composition at the lower boundary of the escape model is constrained by a full photochemical model of the lower atmosphere.  We confirm that molecules dissociate near the 1 $\mu$bar level, and find that complex molecular chemistry does not need to be included above this level.  We also confirm that diffusive separation of the detected species does not occur because the heavy atoms and ions collide frequently with the rapidly escaping H and H$^+$.  This means that the abundance of the heavy atoms and ions in the thermosphere simply depends on the elemental abundances and ionization rates.  We show that, as expected, H and O remain mostly neutral up to at least 3 $R_p$, whereas both C and Si are mostly ionized at significantly lower altitudes.  We also explore the temperature and velocity profiles, and find that the outflow speed and the temperature gradients depend strongly on the assumed heating efficiencies.  Our models predict an upper limit of 8,000 K for the mean (pressure averaged) temperature below 3 $R_p$, with a typical value of 7,000 K based on the average solar XUV flux at 0.047 AU.  We use these temperature limits and the observations to evaluate the role of stellar energy in heating the upper atmosphere.                   
\end{abstract}

\begin{keyword}
Extra-solar planets \sep Aeronomy \sep Atmospheres, composition \sep Photochemistry
\end{keyword}

\end{frontmatter}

\linenumbers

\section{Introduction}
\label{sc:intro}

The detection of hot atomic hydrogen in the upper atmosphere of HD209458b \citep{vidalmadjar03,vidalmadjar04} has inspired numerous attempts to model physical and chemical processes in highly irradiated atmospheres, including rapid escape as one of the most challenging aspects.  Subsequent detection of heavy atoms and ions \citep{vidalmadjar04,linsky10} point out the need for more complex models that include the chemistry associated with these species as well as the collision coupling between them and the major species.  Indeed, close-in extrasolar planets offer a natural laboratory to constrain the theory of rapid escape, including hydrodynamic escape.  This is important because aspects of the theory are controversial, and yet rapid escape is believed to have played a role in shaping the early evolution of the atmospheres in the solar system \citep[e.g.,][]{zahnle86,hunten87}.  Escape may also be a crucial factor in determining atmospheric conditions and habitability of super-Earths and Earth-like planets around M dwarfs \citep[e.g.,][]{tarter07} that may be amenable to spectroscopic studies in the near future \citep[e.g.,][]{charbonneau09}.  

The basic ideas about the nature of the upper atmospheres around close-in EGPs were laid out almost as soon as the first planet, 51 Peg b \citep{mayor95}, was detected.  For instance, \citet{coustenis98} argued that heating by the stellar EUV radiation and interaction with the stellar wind leads to high temperatures of several thousand Kelvins in the upper atmosphere and exosphere of close-in EGPs.  They also suggested that the upper atmosphere is primarily composed of atoms and ions, and that hydrodynamic escape might be able to drag species heavier than H and He into the exosphere.  At the same time, \citet{schneider98} argued that material escaping from the atmospheres of close-in EGPs would form a potentially observable comet-like tail.  When \citet{vidalmadjar03,vidalmadjar04} detected the transits of HD209458b in the stellar FUV emission lines, they also argued that the planet is followed by comet-like tail of escaping hydrogen, and that hydrodynamic escape is required to drag oxygen and carbon atoms to the thermosphere.    

The model of \citet{yelle04,yelle06} was the first attempt to model the aeronomy and escape processes in detail and most of the assumptions in that work have been adopted by subsequent investigators.  It solved the vertical equations of continuity, momentum, and energy for an escaping atmosphere, including photochemistry in the ionosphere and transfer of stellar XUV radiation.  Based on a composition of hydrogen and helium, the results demonstrated that H$_2$ dissociates in the thermosphere, which at high altitudes is dominated by H and H$^+$.  The model also showed that stellar heating leads to temperatures of $\sim$10,000 K in the upper atmosphere, and predicted an energy-limited mass loss rate of 4.7~$\times$~10$^7$ kg~s$^{-1}$ \citep{yelle06}. 

 \citet{yelle04} argued that conditions beyond $\sim$3 $R_p$ were too complex and uncertain to be modeled reliably and therefore chose an upper boundary at 3 $R_p$, rather than at infinity, as adopted in early solar wind models.  This led to a requirement for boundary conditions for the fluid equations at a finite radius.  \citet{yelle04} required consistency between fluid and kinetic simulations, based on the well established fact that kinetic and fluid approaches provide consistent results for the escape flux \citep[e.g.,][]{lemaire73}.  This led to subsonic velocities of a few km~s$^{-1}$ in his model -- although the presence of a sonic point at a higher altitude was not ruled out.     

Many other models for the upper atmospheres of close-in EGPs have been published \citep[e.g.,][]{lammer03,lecavelier04,jaritz05,tian05,erkaev07,garciamunoz07,schneiter07,penz08,holstrom08,murrayclay09,stone09,guo11,trammell11}.  These include one-dimensional, two-dimensional, and three-dimensional models that make different assumptions regarding heating efficiency, the effect of stellar tides, photochemistry, and the escape mechanism.  Despite significant differences in the temperature and velocity profiles, almost all of the existing models agree that close-in EGPs such as HD209458b are surrounded by an extended, hot thermosphere that is undergoing some form of escape.  Most of the models to date concentrate on the distribution of H and H$^+$ in the upper atmosphere.  \citet{garciamunoz07} developed the only model to address the presence of O and C$^+$ in the thermosphere \citep{vidalmadjar04,linsky10}.  This model is otherwise similar to \citet{yelle04}, but it includes the photochemistry of heavy ions, atoms, molecules, and molecular ions.  It also extends to higher altitudes, and includes the effect of substellar tidal forces and stellar wind, albeit in an approximate manner.

\citet{koskinen07a,koskinen07} developed a three-dimensional model for the thermospheres of EGPs at wide orbits.  They pointed out that the atmospheres of close-in EGPs do not escape hydrodynamically unless they receive enough stellar XUV energy to dissociate molecules in the EUV heating layer below the exobase.  Although their results are limited to the specific case of H$_2$, they can be generalized as follows.  The most important molecules H$_2$ (through the formation of H$_3^+$), CO, H$_2$O, and CH$_4$ act as strong infrared coolants in the thermosphere.  High temperatures and rapid escape are only possible once these molecules are dissociated.  \citet{koskinen07} showed that H$_2$ dissociates in the thermosphere of a Jupiter-type planet orbiting a Sun-like star within 0.2 AU.  Once H$_2$ dissociates, it is reasonable to assume that other molecules dissociate too.  At this point the pressure scale height is enhanced by a factor of $\sim$10 when H becomes the dominant species in the thermosphere and temperatures reach 10,000 K.     

It should be noted that a composition of H and H$^+$ with high temperatures does not guarantee that the atmosphere escapes hydrodynamically.  For instance, \citet{koskinen09} showed that hydrodynamic escape is extremely unlikely to occur on a planet such as HD17156b because of its high mass and eccentric orbit.  These types of results have implications on statistical studies that characterize the escape of planetary atmospheres by relying on the so-called energy-limited escape \citep[e.g.,][]{watson81,lecavelier07,sanzforcada10}.  These studies often include an efficiency factor in the mass loss rate that is based on the heating efficiency of the upper atmosphere \citep[e.g.,][]{lammer09}.  Unless the atmosphere is escaping rapidly, the heating efficiency could be considerably larger than the fraction of energy that actually powers escape through adiabatic cooling.  Under diffusion-limited escape or in the Jeans regime the energy-limited escape rate is just an upper limit and the true escape rate can be lower.   

Ideally, the uncertainties in the models can be limited by detailed observations of the escaping species.  At present, multiple observations are only available for HD209458b, and they reveal the presence of H, O, C$^+$, and Si$^{2+}$ at high altitudes in the thermosphere \citep{vidalmadjar03,vidalmadjar04,linsky10}.  Visible and infrared observations have also revealed the presence of Na, H$_2$O, CH$_4$, and CO$_2$ in the lower atmosphere \citep{charbonneau02,knutson08,swain09}.  Taken together, these observations are beginning to reveal the composition and thermal structure in the atmosphere of HD209458b.  The purpose of the current paper is to characterize the density profiles of all of the detected species in the thermosphere, and to explain the presence of the heavy atoms and ions at high altitudes in the upper atmosphere.  The results can be used to infer some basic properties of the atmosphere.                         

To this end, we introduce a one-dimensional escape model for the upper atmosphere of HD209458b that includes the photochemistry of heavy atoms and ions.  As pointed out above, previous models agree broadly on the qualitative nature of the thermosphere but the temperature, density, and velocity profiles predicted by them differ significantly (see Section~\ref{subsc:tempvel}).  Some authors have argued that the density of H in the thermosphere is not sufficient to explain the observed transit depths \citep[see][for a review]{koskinen10}, thus lending support to alternative interpretations of the observations such as the presence of energetic neutral atoms \citep{holstrom08} or a comet-like tail of hydrogen shaped by radiation pressure \citep{vidalmadjar03}.  Accurate modeling of the thermosphere is required to enable better judgment between different explanations of the observations.    

The differencies between previous models arise from different assumptions regarding heating rates and boundary conditions.  In addition to modeling the density profiles of the detected heavy species, we have improved these aspects of the calculations in our work.  For instance, the lower boundary conditions are constrained by results from a detailed photochemical model of the lower atmosphere (Lavvas et al., \textit{in preparation}).  With regard to the upper boundary conditions, we demonstrate that for HD209458b the extrapolated `outflow' boundary conditions \citep[e.g.,][]{tian05} are consistent with recent results from kinetic theory \citep{volkov11a,volkov11b} as long as the upper boundary is at a sufficiently high altitude -- although uncertainties regarding the interaction of the atmosphere with the stellar wind may limit the validity of both boundary conditions.  We highlight the effect of heating efficiency and stellar flux on the density and temperature profiles, and constrain the likely heating rates by using photoelectron heating efficiencies based on the results of \citet{cecchi09} and our own estimates (Section~\ref{subsc:tempvel}).  As a result we provide a robust qualitative description of the density profiles, and constrain the mean temperature and velocity profile in the thermosphere.  A second paper by \citet{koskinen12b} (Paper II) compares our results directly with the observations.  

\section{Methods}
\label{sc:methods}

\subsection{Hydrodynamic model}
\label{subsc:hydromodel}

We use a one-dimensional escape model for HD209458b ($R_p =$~1.32 $R_J$, $M_p =$~0.69 $M_J$, $a =$~0.047 AU) that is similar to the models of \citet{yelle04} and \citet{garciamunoz07}.  Because such models are extensively discussed in the literature, we include only a brief overview of the model here, with the emphasis on how it differs from previous work.  The model solves the one-dimensional equations of motion for an escaping atmosphere composed of several neutral and ionized species:
\begin{eqnarray}
\frac{\partial \rho_s}{\partial t} + \frac{1}{r^2} \frac{\partial}{\partial r} (r^2 \rho_s v)&+&\frac{1}{r^2} \frac{\partial}{\partial r} (r^2 F_s) = \sum_t R_{st} \\
\frac{\partial (\rho v)}{\partial t} + \frac{1}{r^2} \frac{\partial}{\partial r} (r^2 \rho v^2)&=&-\rho g - \frac{\partial p}{\partial r} + f_{\mu} \\
\frac{\partial (\rho E)}{\partial t} + \frac{1}{r^2} \frac{\partial}{\partial r} (r^2 \rho E v)&=&\rho Q_R - p \frac{1}{r^2} \frac{\partial}{\partial r} (r^2 v) \nonumber \\ 
+ \frac{1}{r^2} \frac{\partial}{\partial r} \left( r^2 \kappa \frac{\partial T}{\partial r} \right)&+&\Phi_{\mu}
\end{eqnarray}
where $\rho_s$ is the density of species $s$, $v$ is the vertical velocity, $F_s$ is the diffusive flux of species $s$, $R_{st}$ is the net chemical source term for species $s$, $f_{\mu}$ is a force term arising from viscous acceleration, $E = c_v T$ is the specific internal energy of the gas, $Q_R$ is the specific net radiative heating rate, $\kappa$ is the coefficient of heat conduction, and $\Phi_{\mu}$ is the viscous dissipation functional \citep[e.g.,][]{oneill89}.  The total density and pressure are given by $\rho = \sum_s \rho_s$ and $p = \sum_s n_s k T$, respectively, where electrons contribute to the total pressure.  

We assumed equal temperatures for the neutral species, ions and electrons, and calculated the electron density at each altitude from the requirement of charge neutrality.  The model solves separate continuity equations for each species, but treats the atmosphere otherwise as a single fluid.  The differences in the velocities of the individual species are taken into account by including the diffusive flux $F_s$.  We calculated the fluxes by solving simultaneous equations for multiple species based on the diffusion equation given by \citet{chapman70} (equation 18.2,6, p.344).  We also included a force term due to the ambipolar electric field given by $eE = -(1/n_e) \text{d} p_e / \text{d} r$, where the subscript $e$ refers to electrons, that can be important in highly ionized flows.  The collision terms account for neutral-neutral, resonant and non-resonant ion-neutral, and Coulomb collisions.  This method is in principle similar to those of \citet{yelle04} and \citet{garciamunoz07}.  We verify that the single temperature and diffusion approximations are valid for HD209458b based on our results in Section~\ref{subsc:ionescape}. 

The model includes heat conduction and terms due to viscosity in both the momentum and energy equations.  Thus the equations are consistent with the level of approximation in the Navier-Stokes (NS) equations.  The NS equations themselves are a simplification of the 13-moment solution to the Boltzmann equation \citep[e.g.,][]{gombosi94} that is valid when the Knudsen number $Kn = \Lambda/L << 1$, where $\Lambda$ is the mean free path and $L$ is the typical length scale for significant changes in density or temperature.  Broadly speaking, the equations are valid below the exobase, and terms due to heat conduction and viscosity gain significance as $Kn \rightarrow$~1.  We note that the exobase on HD209458b is typically located at a very high altitude (see Section~\ref{subsc:tempvel}), and viscosity and heat conduction are not particularly important.       

We included species such as H, H$^+$, He, He$^+$, C, C$^+$, O, O$^+$, N, N$^+$, Si, Si$^+$, Si$^{2+}$, and electrons in the hydrodynamic model.  We also generated simulations that included Mg, Mg$^{+}$, Na, Na$^{+}$, K, K$^+$, S, and S$^+$, but the presence of these species did not affect the density profiles of H, O, C$^+$, or Si$^{2+}$ significantly.  The model includes photoionization, thermal ionization, and charge exchange between atoms and ions.  The reaction rate coefficients for these processes are listed in Table~\ref{table:reactions}.  Multiply charged ions were included only if the ionization potential of their parent ion was sufficiently low compared to the thermal energy and radiation field in the upper atmosphere.  We note that our model also includes impact ionization by thermal electrons.  In general, this can be important for species with low ionization potential such as alkali metals \citep[e.g.,][]{batygin10}, although we find photoionization to be more significant in the thermosphere (see Section~\ref{subsc:denprofs}).   

In order to simulate photochemistry in a numerically robust fashion, we coupled the dynamical model with the ASAD chemistry integrator developed at the University of Cambridge \citep{carver97}.  In most cases we used the IMPACT integration scheme that is provided by ASAD.  We did not include any molecules in the present simulations, and thus placed the lower boundary of the hydrodynamic model at $p_0 =$~1 $\mu$bar (see Section~\ref{subsc:photochemistry}).  Molecular chemistry is not significant in the thermosphere, where our results agree qualitatively with \citet{garciamunoz07} despite simpler chemistry (see Section~\ref{subsc:denprofs}).  This is an important result because it implies that complex molecular photochemistry does not need to be included in the models for the thermosphere.  However, the chemistry of molecular ions may be important on HD209458b below the 0.1 $\mu$bar level and it needs to be studied in greater detail.    

\begin{table}[htbp]
 \tiny{
 \centering
 \caption{Reaction rate coefficients}
 \begin{tabular}{@{} lcc @{}}
 \toprule
 \cmidrule( r ){1-3}  
  Reaction  & Rate (cm$^3$ s$^{-1}$) & Reference \\ 
  \midrule
  P1 \ \ \ \  H$\ + \ h\nu \ \rightarrow$~H$^+ \  + \  e$                 & & \citep{hummer63} \\
  P2 \ \ \ \  He$\ + \ h\nu \ \rightarrow$~He$^+ \  + \  e$            & & \citep{yan98} \\
  P3 \ \ \ \  O$\ + \ h\nu \ \rightarrow$~H$^+ \  + \  e$                 & & \citep{verner96} \\
  P4 \ \ \ \  C$\ + \ h\nu \ \rightarrow$~C$^+ \  + \  e$                 & & \citep{verner96} \\
  P5 \ \ \ \  N$\ + \ h\nu \ \rightarrow$~N$^+ \  + \  e$                 & & \citep{verner96} \\
  P6 \ \ \ \  Si$\ + \ h\nu \ \rightarrow$~Si$^+ \  + \  e$               & & \citep{verner96} \\
  P7 \ \ \ \  Si$^+\ + \ h\nu \ \rightarrow$~Si$^{2+} \  + \  e$      & & \citep{verner96} \\
  R1 \ \ \ \  H$^+ + e \rightarrow$~H~$ + \ h\nu$                      & 4.0~$\times$~10$^{-12} (300/T_e)^{0.64}$                                                                                                        & \citep{storey95}  \\
  R2 \ \ \ \  He$^+ + e \rightarrow$~He~$ + \ h\nu$                 & 4.6~$\times$~10$^{-12} (300/T_e)^{0.64}$                                                                                                         & \citep{storey95}  \\
  R3 \ \ \ \  H$\  +\  e \rightarrow$~H$^+ + e + e$                     & 2.91~$\times$~10$^{-8} \left( \frac{1}{0.232+U} \right) U^{0.39} \exp(-U) \ , \ U =  13.6/E_e(eV)$              & \citep{voronov97}  \\
  R4 \ \ \ \  He$\  +\  e \rightarrow$~He$^+ + e + e$                 & 1.75~$\times$~10$^{-8} \left( \frac{1}{0.180+U} \right) U^{0.35} \exp(-U) \ , \ U =  24.6/E_e(eV)$             & \citep{voronov97} \\ 
  R5 \ \ \ \  H$ \ +$~He$^+ \rightarrow$~H$^+ \  + \ $He         & 1.25~$\times$~10$^{-15} (300/T)^{-0.25}$                                                                                                            & \citep{glover07}  \\
  R6 \ \ \ \  H$^+ \ +$~He$\  \rightarrow$~H$ \  + \ $He$^+$  & 1.75~$\times$~10$^{-11} (300/T)^{0.75} \exp(-128,000/T)$                                                                            & \citep{glover07}  \\
  R7 \ \ \ \  O$\  +\  e \rightarrow$~O$^+ + e + e$                     & 3.59~$\times$~10$^{-8} \left( \frac{1}{0.073+U} \right) U^{0.34} \exp(-U) \ , \ U =  13.6/E_e(eV)$              & \citep{voronov97}  \\
  R8 \ \ \ \  C$\  +\  e \rightarrow$~C$^+ + e + e$                     & 6.85~$\times$~10$^{-8} \left( \frac{1}{0.193+U} \right) U^{0.25} \exp(-U) \ , \ U =  11.3/E_e(eV)$              & \citep{voronov97}  \\
  R9 \ \ \ \  O$^+ + e \rightarrow$~O~$ + \ h\nu$                      & 3.25~$\times$~10$^{-12} (300/T_e)^{0.66}$                                                                                                       & \citep{woodall07}  \\
  R10 \ \  C$^+ + e \rightarrow$~C~$ + \ h\nu$                        & 4.67~$\times$~10$^{-12} (300/T_e)^{0.60}$                                                                                                       & \citep{woodall07}  \\
  R11 \ \  C$^+ \ +$~H $\ \rightarrow$~C~$ + \ $ H$^+$        & 6.30~$\times$~10$^{-17} (300/T)^{-1.96} \exp(-170,000/T)$                                                                               & \citep{stancil98}  \\
  R12 \ \  C$ \ +$~H$^+ \ \rightarrow$~C$^+ \ + \ $ H             & 1.31~$\times$~10$^{-15} (300/T)^{-0.213}$                                                                                                            & \citep{stancil98}  \\
  R13 \ \  C$ \ +$~He$^+ \ \rightarrow$~C$^+ \ + \ $ He        & 2.50~$\times$~10$^{-15} (300/T)^{-1.597}$                                                                                                             & \citep{glover07}  \\
  R14 \ \  O$^+ \ +$~H $\ \rightarrow$~O~$ + \ $ H$^+$        & 5.66~$\times$~10$^{-10} (300/T)^{-0.36} \exp(8.6/T)$                                                                                           & \citep{woodall07}  \\
  R15 \ \  O$ \ +$~H$^+ \ \rightarrow$~O$^+ \ + \ $ H             & 7.31~$\times$~10$^{-10} (300/T)^{-0.23} \exp(-226.0/T)$                                                                                     & \citep{woodall07}  \\
  R16 \ \  N$\  +\  e \rightarrow$~N$^+ + e + e$                       & 4.82~$\times$~10$^{-8} \left( \frac{1}{0.0652+U} \right) U^{0.42} \exp(-U) \ , \ U =  14.5/E_e(eV)$               & \citep{voronov97}  \\
  R17 \ \  N$^+ + e \rightarrow$~N~$ + \ h\nu$                        & 3.46~$\times$~10$^{-12} (300/T_e)^{0.608}$                                                                                                       & \citep{aldrovandi73}  \\
  R18 \ \  Si$\  +\  e \rightarrow$~Si$^+ + e + e$                      & 1.88~$\times$~10$^{-7} \left( \frac{1+\sqrt{U}}{0.376+U} \right) U^{0.25} \exp(-U) \ , \ U =  8.2/E_e(eV)$    & \citep{voronov97}  \\
  R19 \ \  Si$^+ + e \rightarrow$~Si~$ + \ h\nu$                       & 4.85~$\times$~10$^{-12} (300/T_e)^{0.60}$                                                                                                          & \citep{aldrovandi73}  \\
  R20 \ \  Si$^+ \  +\  e \rightarrow$~Si$^{2+} + e + e$            & 6.43~$\times$~10$^{-8} \left( \frac{1+\sqrt{U}}{0.632+U} \right) U^{0.25} \exp(-U) \ , \ U =  16.4/E_e(eV)$  & \citep{voronov97}  \\
  R21 \ \  Si$^{2+} + e \rightarrow$~Si$^+$~$ + \ h\nu$                  & 1.57~$\times$~10$^{-11} (300/T_e)^{0.786}$                                                                                                        & \citep{aldrovandi73}  \\
  R22 \ \  H$^+ \ +$~Si $\ \rightarrow$~H~$ + \ $ Si$^+$       & 7.41~$\times$~10$^{-11} (300/T)^{-0.848}$                                                                                                              & \citep{glover07}  \\
  R23 \ \  He$^+ \ +$~Si $\ \rightarrow$~He~$ + \ $ Si$^+$   & 3.30~$\times$~10$^{-9}$                                                                                                                                             & \citep{woodall07}  \\
  R24 \ \  C$^+ \ +$~Si $\ \rightarrow$~C~$ + \ $ Si$^+$        & 2.10~$\times$~10$^{-9}$                                                                                                                                             & \citep{woodall07}  \\
  R25 \ \  H$ \ +$~Si$^{2+} \ \rightarrow$~H$^+ \  + \ $ Si$^+$   & 2.20~$\times$~10$^{-9} (300/T)^{-0.24}$                                                                                                             & \citep{kingdon96}  \\
  R26 \ \  H$^+ \ +$~Si$^+ \ \rightarrow$~H$ \  + \ $ Si$^{2+}$   & 7.37~$\times$~10$^{-10} (300/T)^{-0.24}$                                                                                                           & \citep{kingdon96} \\
  \bottomrule
  \end{tabular}
  \label{table:reactions}}
\end{table}   

The upper atmosphere is heated by stellar XUV radiation.  We simulated heating and photoionization self-consistently by using the model density profiles and the UV spectrum of the average Sun.  The spectrum covers wavelengths between 0.1--3000~\AA.  The XUV spectrum between 0.1--1050~\AA~was generated by the SOLAR2000 model \citep{tobiska00}.  It includes strong emission lines separately and weaker lines binned by 50~\AA.  The Lyman~$\alpha$ line was included with a wavelength spacing of 0.5~\AA~from \citet{lemaire05} and the rest of the spectrum was taken from \citet{woods02}.  We assumed that most of the Lyman $\alpha$ radiation absorbed by H is resonantly scattered and does not contribute significantly to the heating of the atmosphere.  This is because the lifetime of the 2p state of H is only 1.6 ns, compared with the typical collision timescale of $\sim$1 s near the temperature peak in the thermosphere of HD209458b.      

References for photoabsorption cross sections of the different species are included in Table~\ref{table:reactions}.  In general, we divided the incident stellar flux by a factor of 4 to account for uniform redistribution of energy around the planet.  This is expected to be approximately valid in the lower thermosphere based on the three-dimensional simulations of \citet{koskinen10b}.  In the extended upper thermosphere, on the other hand, radiation passes through to the night side and leaves only a small region free of direct heating \citep[e.g., see Figure 2 of][]{koskinen07}.  The current model also includes heating due to photoabsorption by C, O, N, and metals.  This is relatively insignificant -- although it leads to some additional heating in the lower thermosphere by FUV radiation.

Heating of the thermosphere is mostly driven by photoionization and the generation of photoelectrons, although direct excitation of atoms and molecules may also play a role.  Photoelectrons excite, ionize, and dissociate atoms and molecules until they lose enough energy and become thermalized i.e., share their energy with thermal electrons in Coulomb collisions.  Thermal electrons share their energy with ions and eventually, the neutral atmosphere.  In highly ionized atmospheres such as on HD209458b the photoelectron heating efficiency can be close to 100~\% \citep{cecchi09}, depending on the energy of the photoelectrons.  We used scaled heating efficiencies that depend on photoelectron energy to estimate the net heating efficiency in the atmosphere (Section~\ref{subsc:tempvel}).  

Generally, we refer to two different definitions of heating efficiency in Section~\ref{subsc:tempvel} in order to highlight the effect of heating efficiency on the temperature and velocity profiles.  The net heating efficiency $\eta_{\text{net}}$ is defined simply as the fraction of the absorbed stellar energy that heats the atmosphere.  Photoelectron heating efficiency, on the other hand, applies to photoelectrons with energy $E_p = h \nu - I_s$ where $I_s$ is the ionization potential of species $s$ and $h \nu$ is the energy of the ionizing photon.  The photoelectron heating efficiency is the fraction of $E_p$ that heats the thermosphere, and it is generally higher than $\eta_{\text{net}}$ because it does not account for recombination.  The net heating efficiency is often used to calculate mass loss rates for extrasolar planets \citep[e.g.,][]{lammer09}.  Therefore it is important not to confuse these two definitions of heating efficiency.  We included radiative cooling by recombination under the assumption that the thermosphere is optically thin to the emitted photons.  This implies that the ionization potential energy $I_s$ never contributes to heating at any levels.  We also considered the influence of Lyman~$\alpha$ cooling by excited H, although this cooling rate is uncertain and likely to be low for HD209458b.  We discuss the effect of different cooling rates further in Section~\ref{subsc:tempvel}.
 
\subsubsection{Lower boundary conditions}  
\label{subsc:photochemistry}      

As stated above, we placed the lower boundary of the hydrodynamic model at $p_0 = $~1 $\mu$bar and did not include H$_2$ or other molecules in the model.  This decision was motivated by photochemical calculations for HD209458b (Lavvas et al., \textit{in preparation}) that we used to constrain the lower boundary condition.  The photochemical model was originally developed for the atmosphere of Titan \citep{lavvas08a,lavvas08b} but it was recently expanded to simulate EGP atmospheres.  It calculates the chemical composition from the deep troposphere (1000 bar) up to the thermosphere above the 0.1 nbar level by solving the coupled continuity equations for all species based on a database of $\sim$1,500 reaction rate coefficients and 103 photolysis processes.  Forward and reverse rates are included for each reaction with the ratio of the rate coefficients defined by thermochemical data.  Thus, the results are consistent with thermochemical equilibrium at deep atmospheric levels but differences develop at higher altitudes due to photolysis, diffusion, and eddy mixing.  At the lower boundary the chemical abundances of the main species (H, C, N, and O) are set to their thermodynamic equilibrium values and, depending on the vertical temperature profile and their abundances, species are allowed to condense.  

\begin{figure}
  \centering
  \includegraphics[width=0.7\textwidth]{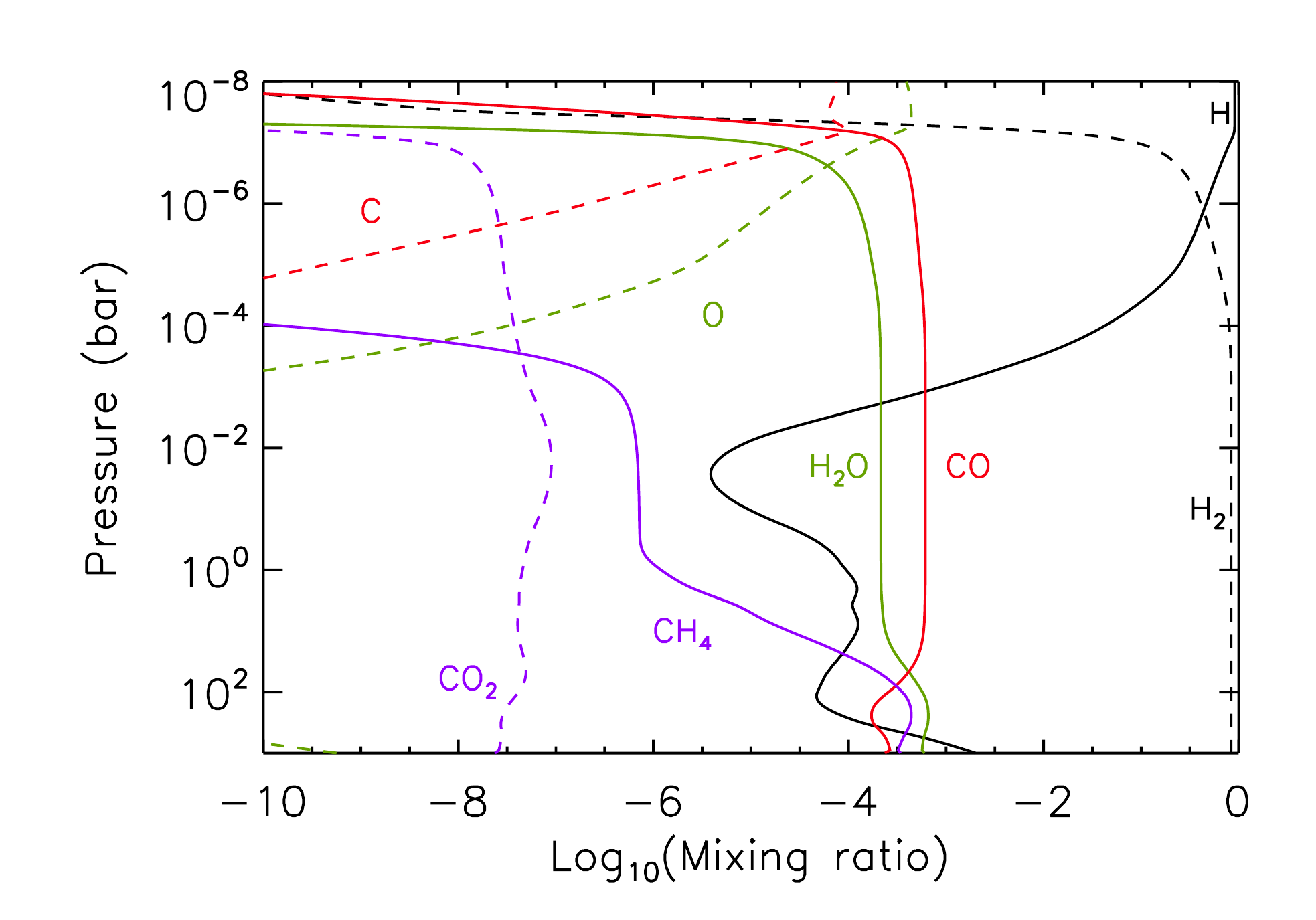}
  \caption{Mixing ratios of key oxygen and carbon-bearing species in the dayside atmosphere of HD209458b (Lavvas et al., \textit{in preparation}).}
  \label{fig:photochemistry}
\end{figure}

Figure~\ref{fig:photochemistry} shows the mixing ratios of H$_2$, H, H$_2$O, O, CH$_4$, CO, CO$_2$, and C as a function of altitude from the photochemical model.  In general, the results are similar to those of \citet{moses11}.  The H$_2$/H transition is located near 1 $\mu$bar.  At lower pressures, the mixing ratio of H$_2$ decreases rapidly with altitude and falls below 0.1 above the 0.1~$\mu$bar level.  In agreement with \citet{garciamunoz07}, the dissociation of H$_2$ is caused by dissociation of H$_2$O, which leads to the production of OH radicals that attack the H$_2$ molecules.  We note that the exact location of the H$_2$/H transition depends on the temperature profile and, depending on the thermal structure, it could occur even below the 1 $\mu$bar level.

The major oxygen-bearing molecules, CO and H$_2$O, have roughly equal abundances from 10 to 10$^{-5}$ bar.  This is in line with thermochemical equilibrium at the temperatures and pressures relevant to HD209458b \citep{lodders02}.  H$_2$O and CO are effectively dissociated at pressures lower than 0.3 and 0.1 $\mu$bar, respectively.  We note that molecular abundances could be significant at 0.1--1 $\mu$bar, and technically the results of the hydrodynamic calculations are only valid above the 0.1 $\mu$bar level.  The presence of molecules such as H$_2$, H$_2$O, and CO can lead to enhanced UV heating as well as efficient radiative cooling by H$_3^+$, H$_2$O and CO in the 0.1--1 $\mu$bar region.

The photochemical model also includes the chemistry of silicon.  Due to condensation into forsterite and enstatite \citep[e.g][]{visscher10}, the abundance of Si in the observable atmospheres of giant planets was thought to be negligible and thus the photochemistry of silicon in planetary atmospheres has not been studied before.  The photochemical calculations indicate that, in agreement with thermochemical calculations \citep{visscher10}, SiO is the dominant silicon-bearing gas.  SiO dissociates in the thermosphere at a similar pressure level as H$_2$O and CO.  We note that the detection of Si$^{2+}$ in the thermosphere implies that silicon does not condense in the atmosphere of HD209458b (Paper II).  

In addition to composition, lower boundary conditions are required for temperature and velocity.  We specified $T_0$ and $p_0$ at the lower boundary, and used them to calculate $\rho_0$ from the ideal gas law.  The steady state continuity equation $\rho_0 v_0 r^2 = F_c$, where $F_c$ is the flux constant, was used to calculate the velocity $v_0$ at the lower boundary during each time step.  The flux constant is solved self-consistently by the model, and it depends largely on the terms in the energy equation.  In general we assumed that $T_0 \approx$~1,300 K, which is close to the effective temperature of the planet.  We note that this temperature is largely unconstrained.  Radiative transfer models for close-in EGPs \citep[e.g.,][and references therein]{showman09} do not account for heating by stellar UV radiation or possible opacity sources created by photochemistry \citep[e.g.,][]{zahnle09}.  Therefore these models may not accurately predict the temperature at the base of the thermosphere.   

\citet{sing08a,sing08b} used observations of the Na D line profile to constrain the temperature profile in the upper atmosphere.  They suggested that Na condenses into clouds near the 3 mbar level, and thus predicted a deep minimum in temperature in this region that is required for condensation.  The detection of Si$^{2+}$ implies that condensation of Na in the lower atmosphere is unlikely (Paper II), and thus this result is unreliable.  Relying on the same observations, \citet{vidalmadjar11a,vidalmadjar11b} predicted that the temperature increases steeply from 1,300 K to 3,500 K near the 1 $\mu$bar level.  However, their method to retrieve the temperature relies on the density scale height of Na to express the optical depth along the line of sight (LOS).  This is not consistent with the argument that Na is depleted above the 3 mbar level.  If such a depletion takes place, the density scale height of Na is not the same as the scale height of the atmosphere and it cannot be used to retrieve temperatures.  

\subsubsection{Upper boundary conditions}
\label{subsc:upbound}

Previous models of the thermosphere disagree on the details of the density and temperature profiles \citep[e.g.,][]{yelle04,tian05,garciamunoz07,murrayclay09}.  This is partly due to different boundary conditions, although assumptions regarding the heating rates and composition are probably more important (see Section~\ref{subsc:tempvel}).  Unfortunately, the planetary wind equations can have an infinite number of both subsonic and supersonic solutions.  In time-dependent models, the upper boundary conditions in particular can determine if the solution is subsonic or supersonic, and they can alter the temperature and velocity profiles \citep[e.g.,][]{garciamunoz07}.  The choice of proper boundary conditions is therefore important.

\citet{volkov11a,volkov11b} studied the escape of neutral atmospheres under different circumstances by using the kinetic Monte Carlo (DSMC) method.  Because the fluid equations are a simplification of the kinetic equations at low values of $Kn$, the hydrodynamic model should ideally be consistent with the DSMC results both above and below the exobase.  \citet{volkov11a,volkov11b} found that the nature of the solutions depends on the thermal escape parameter $X_0 = G M_p m / k T_0 r_0$ and the Knudsen number $Kn_0$ at the lower boundary $r_0$ of a region where diabatic heating is negligible.  They argued that hydrodynamic escape is possible when $X_0 <$~2--3 \citep[see also][]{opik63,hunten73}.  When $X_0 \gtrsim$~3, on the other hand, the sonic point is at such a high altitude that the solution is practically subsonic and with $X_0 \gtrsim$~6 the escape rate is similar to the thermal Jeans escape rate.  

The results of the DSMC calculations can be incorporated into hydrodynamic models with appropriate upper boundary conditions.  \citet{volkov11a,volkov11b} suggest that the modified Jeans escape rate, which is based on the drifting Maxwellian velocity distribution function, is a good approximation of the DSMC results in fluid models, consistent with \citet{yelle04}.  In order to contrast the modified Jeans upper boundary conditions (hereafter, the modified Jeans conditions) with other possibilities, we used them and the extrapolated upper boundary conditions (hereafter, the `outflow' conditions) adopted by \citet{tian05} and \citet{garciamunoz07} in our simulations.  In general, we placed the upper boundary at 16 R$_p$.  The impact of the boundary conditions is discussed in Section~\ref{subsc:boundaries}.  

We formulated the outflow conditions simply by extrapolating values for density, temperature and velocity with a constant slope from below.  For the modified Jeans conditions, we calculated the effusion velocity $v_s$ at the upper boundary separately for each species by using equation (9) from \citet{volkov11b}.  Using this equation violates the conservation of electric charge at the upper boundary because the small mass of the electrons causes their velocity to be much larger than the velocity of the protons.  In reality charge separation is prevented by the generation of an ambipolar electric field that ensures that the vertical current is zero at the upper boundary.  This electric field causes the ions to escape faster while it slows the electrons down.  Effectively this lowers the escape velocity ($v_{\text{esc}} = \sqrt{2 GM/r}$) of the ions and increases the escape velocity of the electrons.  

In order to incorporate the ambipolar electric field in the modified Jeans conditions we expressed the Jeans parameters for ions and electrons as:
\begin{eqnarray}
X_i&=&\frac{GMm_i}{kTr} - \frac{\phi_e q_i}{k T} \\  
X_e&=&\frac{GMm_e}{kTr} + \frac{\phi_e |q_e|}{k T}
\end{eqnarray}
where $\phi_e$ is the ambipolar electric potential, $q_{i,e}$ is the electric charge and subscripts $i$ and $e$ stand for electrons and ions, respectively.  We used these Jeans parameters to calculate the effusion velocities for the electrons and ions, and then solved iteratively for $\phi_e$ by using the condition of zero current i.e., $n_e |q_e| v_e = \sum_i n_i q_i v_i$.  This approach is consistent with kinetic models for the solar and polar winds \citep{lemaire71a,lemaire71b}.  Having obtained the correct effusion velocities for the charged and neutral species, we evaluated the mass weighted outflow velocity from:
\begin{equation}
v = \frac{1}{\rho} \sum_s \rho_s v_s 
\end{equation}
and used this velocity as an upper boundary condition.  The values of temperature and density that are required for this calculation were extrapolated from below.  As the model approaches steady state, the solution approaches a value of $v$ at the upper boundary that is consistent with the modified Jeans velocity.            

\subsubsection{Numerical methods}

We solved the equations of motion on a grid of 400--550 levels with increasing altitude spacing.  The radius $r_n$ from the center of the planet at level $n$ is thus given by a geometric series \citep[e.g.,][]{garciamunoz07}:
\begin{equation}
r_n =  r_1 + \sum_{i=1}^{n-1} f_c^i \delta z_0 
\end{equation}  
where $r_1 =$~1.08 R$_p$, $\delta z_0 =$~10 km, and $f_c =$~1.014.  We solved the equations of motions in two parts, separating advection (Eulerian terms) from the other (Lagrangian) terms.  The Lagrangian solution is performed first, and all variables are updated before advection.  We used the flux conservative van Leer scheme \citep[e.g.,][]{vanleer79} for advection, and the semi-implicit Crank-Nicholson scheme \citep[e.g,][]{jacobson99} to solve for viscosity and conduction in the momentum and energy equations, respectively.  We employed a time step of 1 s in all of our calculations.  Despite the sophisticated numerical apparatus, the model is still occasionally unstable, particularly in the early stages of new simulations.  The primary source of the instabilities are pressure fluctuations (sound waves) that are not balanced by gravity.  We used a two-step Shapiro filter \citep{shapiro70} periodically to remove numerical instabilities.  We assumed that a steady state has been reached once the flux constant $F_c$ is constant with altitude and the flux of energy is approximately conserved.

\section{Results}
\label{sc:results}   

\subsection{Temperature and velocity profiles}
\label{subsc:tempvel}

\begin{figure}
  \centering
  \includegraphics[width=0.7\textwidth]{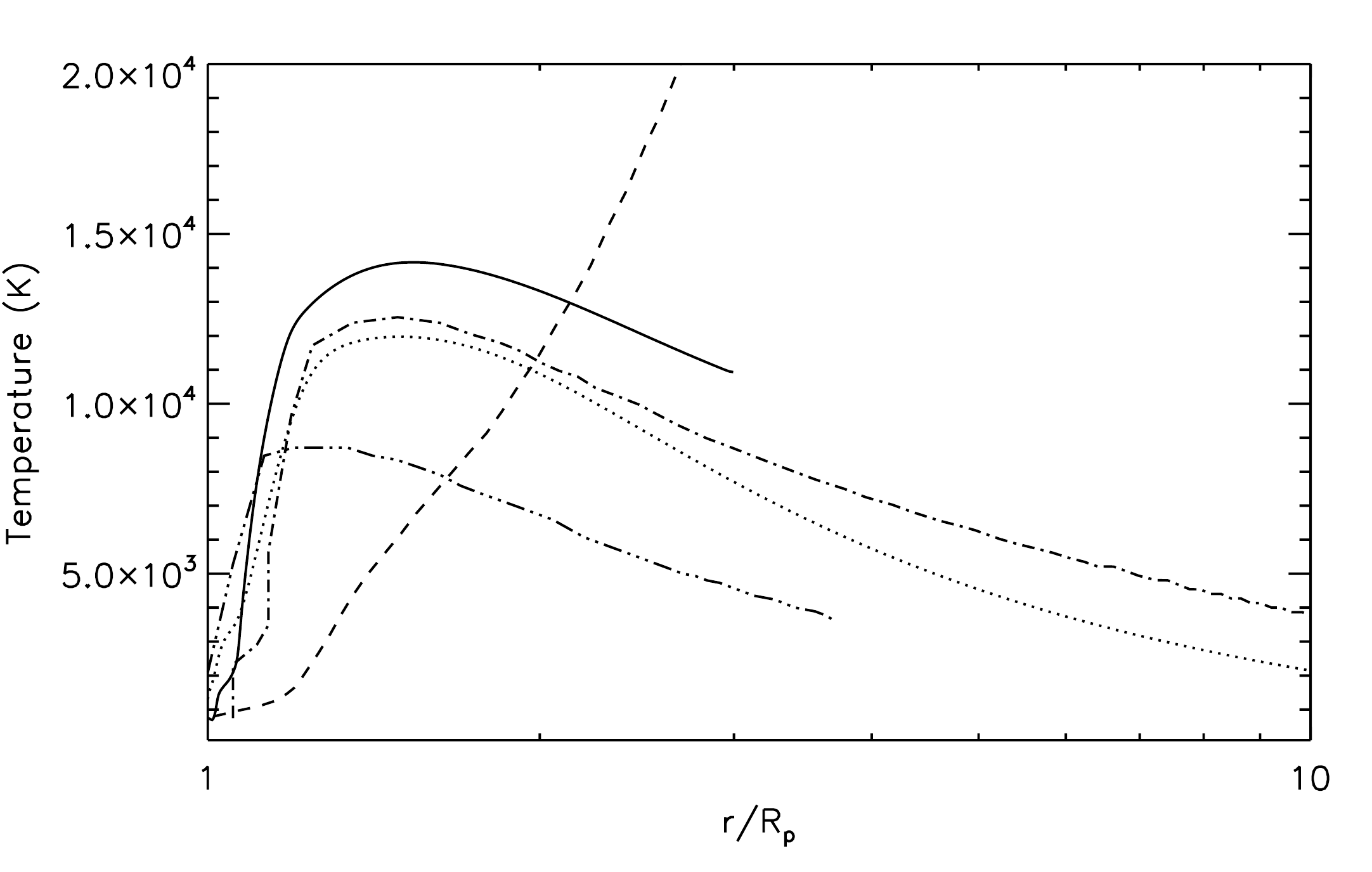}
  \caption{Some examples of temperature profiles from earlier models of the upper atmosphere of HD209458b.  The solid line is from Figure 1 of \citet{yelle04}, the dotted line is from the C2 model in this work (see Section~\ref{subsc:cecchieff}), the dashed line is the atomic hydrogen model from Figure 11 of \citet{tian05}, the dashed-dotted line is the SP model from Figure 3 of \citet{garciamunoz07}, and the dashed-triple-dotted line is from Figure 1 of \citet{murrayclay09}}.
  \label{fig:modelcomp}
\end{figure}

In this section we constrain the range of mean temperatures and velocities based on stellar heating in the thermosphere of HD209458b and similar close-in EGPs.  We discuss the general dependency of the temperature and velocity profiles on the net heating efficiency and stellar flux, and relate this discussion to new temperature and velocity profiles for HD209458b that are based on realistic photoelectron heating efficiencies calculated specifically for close-in EGPs.  This discussion is necessary because the temperature and velocity profiles from previous models of the upper atmosphere differ significantly, and the differences affect the density profiles of the observed species (see Section~\ref{subsc:denprofs}).  As an example, Figure~\ref{fig:modelcomp} shows the temperature profiles based on several earlier models.  In addition to boundary conditions, the discrepancies evident in this figure arise from different assumptions about the heating rates.

\subsubsection{General dependency}
\label{subsc:gendep}

\begin{figure}
  \centering
  \includegraphics[width=0.7\textwidth]{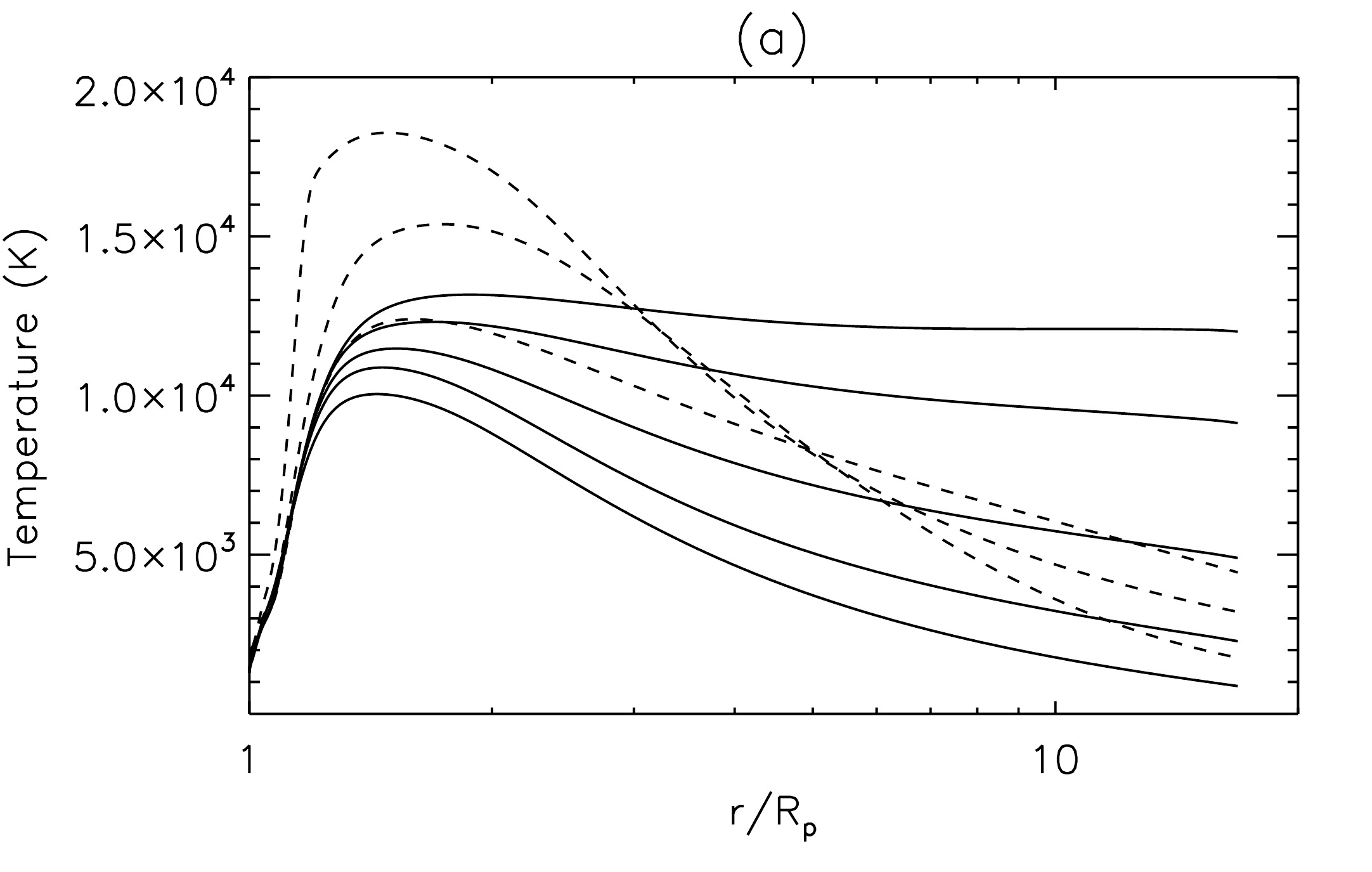}
  \includegraphics[width=0.7\textwidth]{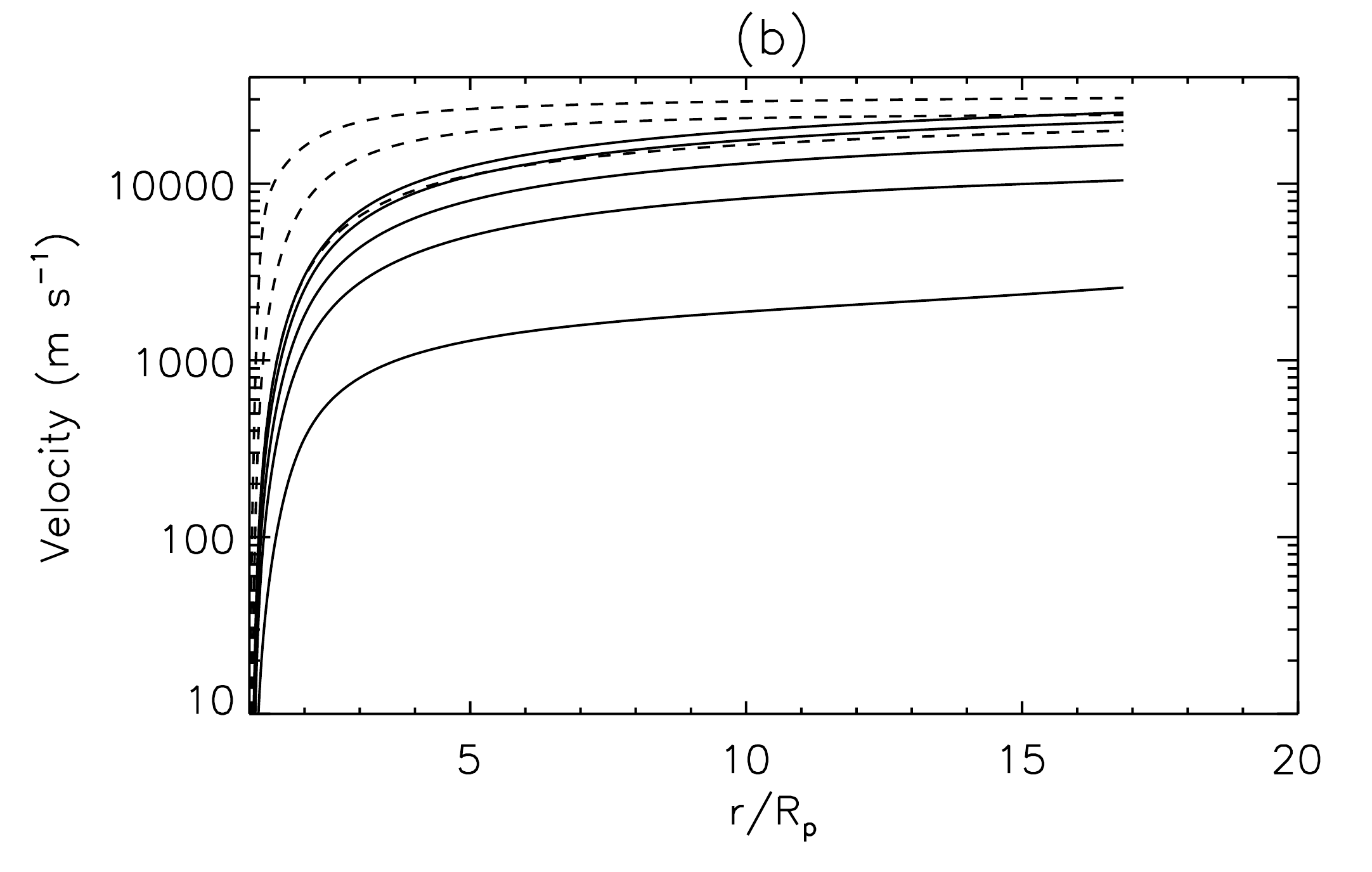}
  \caption{Temperature (a) and velocity (b) profiles in the upper atmosphere of HD209458b based on different heating efficiencies and stellar XUV fluxes.  The solid lines show models based on the average solar flux with $\eta_{\text{net}}$ of 0.1, 0.3, 0.5, 0.8, and 1 (from bottom to top).  The dashed lines show models with $\eta_{\text{net}} =$~0.5 and stellar flux of 2x, 10x, and 100x the solar average flux (in order of increasing peak temperature).}
  \label{fig:gentempvel}
\end{figure}

We note that the thermal structure in the upper atmospheres of the giant planets in the solar system is not very well understood despite modeling and observations that are far more extensive than those available for extrasolar planets \citep[e.g.,][]{yelle04b}.  It is therefore useful test the reaction of the model to different heating rates and profiles.  We used our model to calculate temperature and velocity profiles based on different heating efficiencies and stellar fluxes.  These profiles are shown in Figure~\ref{fig:gentempvel}.  First, we used the average solar spectrum (Section~\ref{subsc:hydromodel}) and varied the net heating efficiency $\eta_{\text{net}}$ from 0.1 to 1.  Second, we multiplied the solar spectrum by factors of 2x, 10x, and 100x, and used $\eta_{\text{net}} =$~0.5.  The range of enhanced fluxes covers solar maximum conditions and stars that are more active than the sun.  In each case we assumed that $\eta_{\text{net}}$ is constant and does not vary with altitude.  We used planetary parameters of HD209458b and set the upper boundary to 16 $R_p$ with outflow boundary conditions, and the lower boundary to the 1 $\mu$bar level with a temperature of 1,300 K.   

A net heating efficiency of 50 \% is similar to the heating efficiency in the Jovian thermosphere \citep{waite83}, and we may consider this as a representative case of a typical gas giant (hereafter, the H50 model).  The maximum temperature in the H50 model is 11,500 K and the temperature peak is located near 1.5 $R_p$ ($p = 0.3$ nbar).  As $\eta_{\text{net}}$ varies from 0.1 to 1, the peak shifts from 1.4 $R_p$ (0.5 nbar) to 1.9 $R_p$ (0.1 nbar) and the maximum temperature varies from 10,000 K to 13,200 K.  It is interesting to note that the temperature profile depends strongly on the heating efficiency but the location of the peak and the maximum temperature depend only weakly on $\eta_{\text{net}}$.  This is because the vertical velocity increases with heating efficiency, leading to more efficient cooling by faster expansion that controls the peak temperature while enhanced advection and high altitude heating flatten the temperature gradient above the peak.  As a result, the temperature profile is almost isothermal when $\eta_{\text{net}} =$~1.

It is also interesting that the temperature profiles in the models that are based on $\eta_{\text{net}} =$~0.5 and the solar flux enhanced by factors of 2--100 differ from models with the average solar flux and a higher heating efficiency.  For instance, one might naively assume that a model with $\eta_{\text{net}} =$~0.5 and 2x the average solar flux would be similar to a model with the average solar flux and $\eta_{\text{eta}} =$~1.  Suprisingly, this is not the case -- despite the fact that the mass loss rates from these models are identical.  This is because of the way radiation penetrates into the atmosphere -- doubling the incoming flux is not the same as doubling the heating rate at each altitude for the same flux.  As the stellar flux increases further, the temperature peak shifts first to higher altitudes, and then to lower altitudes so that for the 100x case the peak is located again near 1.5 $R_p$.  Despite the hugely increased stellar flux, the peak temperature is only 18,300 K for the 100x case.  This is again because the enhanced adiabatic and advective cooling driven by faster expansion control the temperature even in the absence of efficient radiative cooling mechanisms.     

\citet{koskinen10} suggested that the mean temperature of the thermosphere below 3 $R_p$ can be constrained by observations, and used their empirical model to fit temperatures to the H Lyman~$\alpha$ transit data \citep{vidalmadjar03,benjaffel07,benjaffel08}.  A quantity that can be compared with their results is the pressure averaged temperature of the hydrodynamic model, which is given by:
\begin{equation}
\overline{T_p} = \frac{\int_{p_1}^{p_2} T( p ) \  \text{d} (\ln p)}{\ln (p_2/p_1)}.
\label{eqn:meantemp}
\end{equation}  
For $\eta_{\text{net}}$ ranging from 0.1 to 1, the pressure averaged temperature below 3 $R_p$ varies from 6,200 K to 7,800 K for the average solar flux.  In the H50 model the pressure averaged temperature is 7,000 K.  We note that $\overline{T}_p$ is a fairly stable feature of our solutions -- in contrast to the details of the temperature profile and velocities it is relatively insensitive to different assumptions about the boundary conditions and heating efficiencies.  Obviously, $\overline{T}_p$ depends on the stellar flux, although it only increases to 9,800 K in the 100x case.  

\citet{koskinen10} inferred a mean temperature of 8,250 K in the thermosphere of HD209458b with $p_0 =$~1 $\mu$bar (the M7 model).  Taken together with our results based on solar XUV fluxes, this implies a relatively high heating efficiency.  Alternatively, with $\eta_{\text{net}} =$~0.5 it may imply that the XUV flux of HD209458b is 5--10 times higher than the corresponding solar flux.  This type of an enhancement is not impossible.  The activity level of the star depends on its rotation rate, and the rotation rate of HD209458 may be twice as fast as the rotation rate of the sun \citep{silvavalio08}.  However, the uncertainty of the observed H Lyman~$\alpha$ transit depths accommodates a range of temperatures, and thus we are unable to derive firm constraints on the heating rates from the observations.  In general, though, the pressure averaged temperature provides a useful connection between observations and temperatures predicted by models that can be exploited to constrain heating rates.

The effect of changing the heating efficiency on the velocity profile is quite dramatic.  As $\eta_{\text{net}}$ ranges from 0.1 to 1 (with the average solar flux), the velocity at the upper boundary increases from 2.6 km~s$^{-1}$ to 25 km~s$^{-1}$.  However, the velocity does not increase linearly with stellar flux or without a bound -- in the 100x case the velocity at the upper boundary is only 30 km~s$^{-1}$.  An interesting qualitative feature of the solutions is that the sonic point moves to a lower altitude with increasing heating efficiency or stellar flux.  With $\eta_{\text{net}} =$~0.1 the isothermal sonic point is located above the upper boundary whereas with $\eta_{\text{net}} =$~1 it is located at 4 $R_p$.  This behavior is related to the temperature gradient and it is discussed further in Section~\ref{subsc:sonicpoint}.  Basically the sonic point, when it exists, moves further from the planet as the high altitude heating rate decreases.  

It is now clear that assumptions regarding the heating efficiency and radiative transfer have a large impact on the temperature and velocity profiles and the results from the previous models reflect this fact (see Figure~\ref{fig:modelcomp}).  The differences between models have implications on the interpretation of observations.  For instance, \citet{vidalmadjar03} and \citet{linsky10} suggested that the UV transit observations probe the velocity structure of the escaping gas.  Obviously, the nature of this velocity structure depends on the properties of the upper atmosphere.  On the other hand, \citet{benjaffel10} argued that the observations point to a presence of hot energetic atoms and ions within the Roche lobe of the planet.  We believe that it is important to properly quantify the role of stellar heating in creating the hot, escaping material before other options are pursued.  This means that detailed thermal structure calculations that rely on a proper description of photoelectron heating efficiencies are required.  Below we discuss a new approach to modeling the temperature profile in the thermosphere of HD209458b and its impact on the velocity and density profiles.

\subsubsection{Energy balance and temperatures on HD209458b}
\label{subsc:cecchieff}

\begin{figure}
  \centering
  \includegraphics[width=0.7\textwidth]{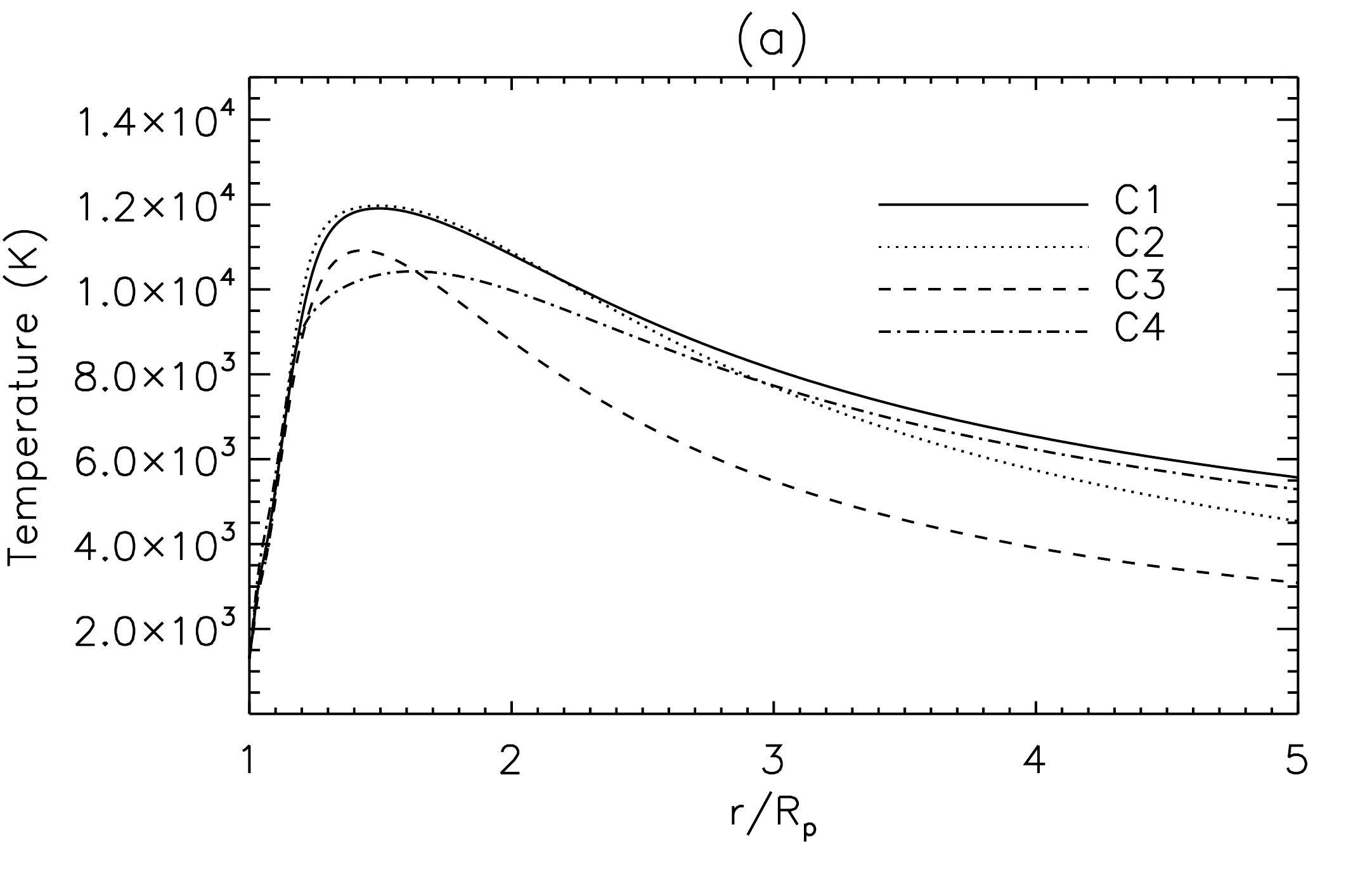}
  \includegraphics[width=0.7\textwidth]{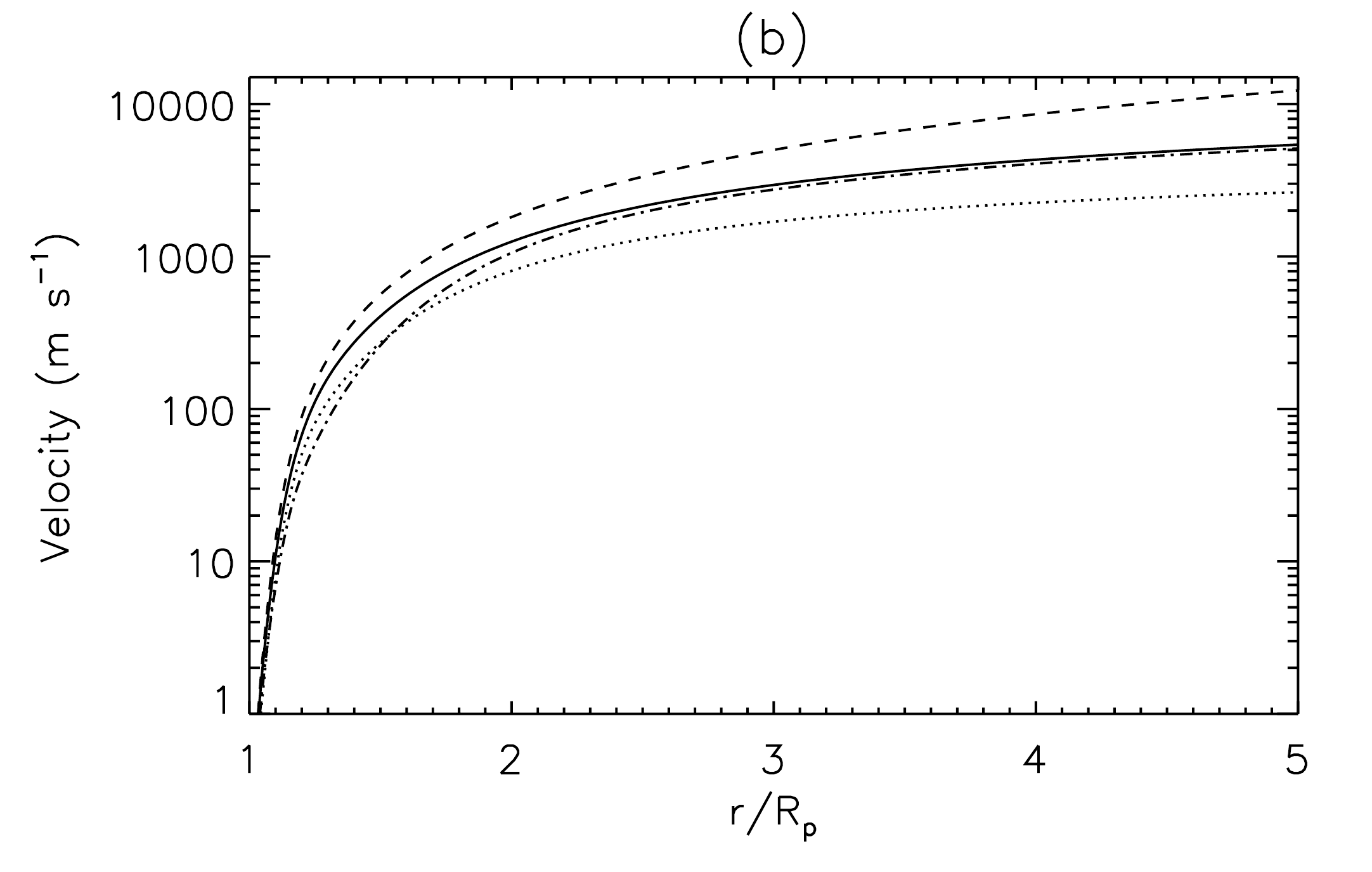}
  \caption{Temperature (a) and velocity (b) profiles in the upper atmosphere of HD209458b based on different models (see Table~\ref{table:models} for the input parameters).}
  \label{fig:tempvel}
\end{figure}

In the previous section we discussed models where the net heating efficiency $\eta_{\text{net}}$ was fixed at a constant value at all altitudes.  In this section we discuss more realistic models of HD209458b that rely on new approximations of photoelectron heating efficiency and derive an estimate of $\eta_{\text{net}}$ based on these models.  Here we also include radiative cooling from recombination and, in one case, H Lyman~$\alpha$ emissions by excited H \citep{murrayclay09}.  Our aim was to calculate the most likely range of temperatures in the thermosphere of HD209458b based on average solar fluxes.  Figure~\ref{fig:tempvel} shows the temperature and velocity profiles at 1--5 $R_p$ based on different approximations (see Table~\ref{table:models} for the input parameters).  Model C1 assumes a constant photoelectron heating efficiency of 93 \% at all altitudes and photoelectron energies.  This heating efficiency is appropriate for photoelectrons created by 50 eV photons at an electron mixing ratio of $x_e =$~0.1 \citep{cecchi09}.  Model C2 is otherwise similar to C1 but the heating efficiency varies with photoelectron energy and altitude (see below).  Models C3 and C4 are also based on C1, but C3 includes the substellar tidal forces in the equations of motion \citep[e.g.,][]{garciamunoz07} and C4 includes Lyman $\alpha$ cooling.  All of these models are based on the outflow boundary conditions for temperature, velocity, and density.      

\begin{table}[htbp]
 \centering
 \caption{Model input parameters and results}
 \begin{tabular}{@{} ccccc @{}}
 \toprule
 \cmidrule( r ){1-5}
  Model$^{\textrm{a}}$  & $r_{\infty}$ (R$_p$)$^{\textrm{b}}$  & $\eta_{\textrm{net}}$$^{\textrm{c,d}}$ & $\dot{M}$ (10$^7$ kg~s$^{-1}$)  & $\overline{T}_{p}$ (K)$^{\textrm{e}}$ \\
 \midrule 
  C1  &  16 E     &  0.56 C  &  5.6  & 7250  \\
  C2  &  16 E     &  0.44 V  &  4.0  & 7200  \\
  C3  &  16 E,T  &  0.57 C &  6.4  & 6450  \\
  C4  &  16 E     &  0.46 C &  4.5  & 7110  \\
  C5  &  36 J      &  0.56 C &  5.6  & 7290  \\
  C6  &  16 J      &  0.45 V &  3.9  & 7310 \\
 \bottomrule 
 \end{tabular}
 \caption*{\small{$^a$All models assume $p_0 =$~10$^{-6}$ bar and $T_0 =$~1,300 K. \\ 
                  $^b$E - Outflow conditions, J - Modified Jeans conditions, T - Substellar tide. \\ 
                  $^c$Net heating efficiency (see Section~\ref{subsc:hydromodel}) i.e., the ratio of the net heating flux at all wavelengths to the unattenuated stellar flux (0.45 W~m$^{-2}$) at wavelengths shorter than 912~\AA. \\
                  $^d$C - Constant photoelectron heating efficiency, V - Varying photoelectron heating efficiency (see text). \\ 
                  $^e$Pressure averaged temperature below 3 $R_p$.}}
  \label{table:models}
\end{table}

\citet{cecchi09} also estimated the heating efficiencies for photoelectrons released by photons of energy $E \gtrsim$~50 eV at different values of the electron mixing ratio $x_e$.  We used their heating efficiencies for $x_e =$~0.1 in the C2 model.  They parameterized their results in terms of the vertical column density $N_{\textrm{H}}$ of H.  We fitted the heating efficiency as a function of $N_{\textrm{H}}$ for 50 eV photons with a regular transmission function, and modified this function accordingly for different cutoff altitudes and heating efficiencies of photons with different energies [see Figures 3 and 4 of \citet{cecchi09}].  We note that $x_e \approx$~0.1 near the temperature peak of our models and thus the results are appropriate for our purposes.  However, they are only applicable to photons with $E \gtrsim$~50 eV.  We used simple scaling to estimate the heating efficiencies for low energy photons with $E <$~50 eV.     

As $N_{\textrm{H}}$ increases, the heating efficiency for 50 eV photons saturates at 93 \%.  We assumed that the saturation heating efficiency for low energy photons is also 93 \%.  In reality, this heating efficiency may be closer to 100 \% but the difference is small.  In order to estimate the altitude dependence of the heating efficiency, we note that the rate of energy deposition by Coulomb collisions between photoelectrons of energy $E_p$ and thermal electrons with a temperature $T$ can be estimated from:
\begin{equation}
- \frac{\textrm{d} F_{E}}{\textrm{d} r} = L(E_p, e) \Phi_{pe} n_e  \ \ \ [\textrm{eV} \ \textrm{cm}^{-3} \ \textrm{s}^{-1}]
\end{equation}
where $F_{E}$ is the flux of energy, $\Phi_{pe}$ is the photoelectron flux (cm$^{-2}$~s$^{-1}$), $n_e$ is the density of thermal electrons (cm$^{-3}$) and
\[ L(E_p, e) = \frac{3.37 \times 10^{-12}}{n_e^{0.03} E_p^{0.94}} \left( \frac{E_p - E_e}{E_p - 0.53 E_e} \right)^{2.36}  \ \ \ [\textrm{eV} \ \textrm{cm}^2], \] 
with $E_e =$~8.618~$\times$~10$^{-5} T_e$ is the stopping power \citep{swartz71}\footnote{Due to a historical precedent, the units here are in cgs.}.  Assuming that all of the energy is deposited by electrons that are thermalized within a path element $dr$, we can estimate the e-folding length scale for thermalization of photoelectrons with different energies as follows:
\begin{equation}
\Lambda_{pe} \approx \frac{E_p}{n_e L}. 
\end{equation}  

\begin{figure}
  \centering
  \includegraphics[width=0.7\textwidth]{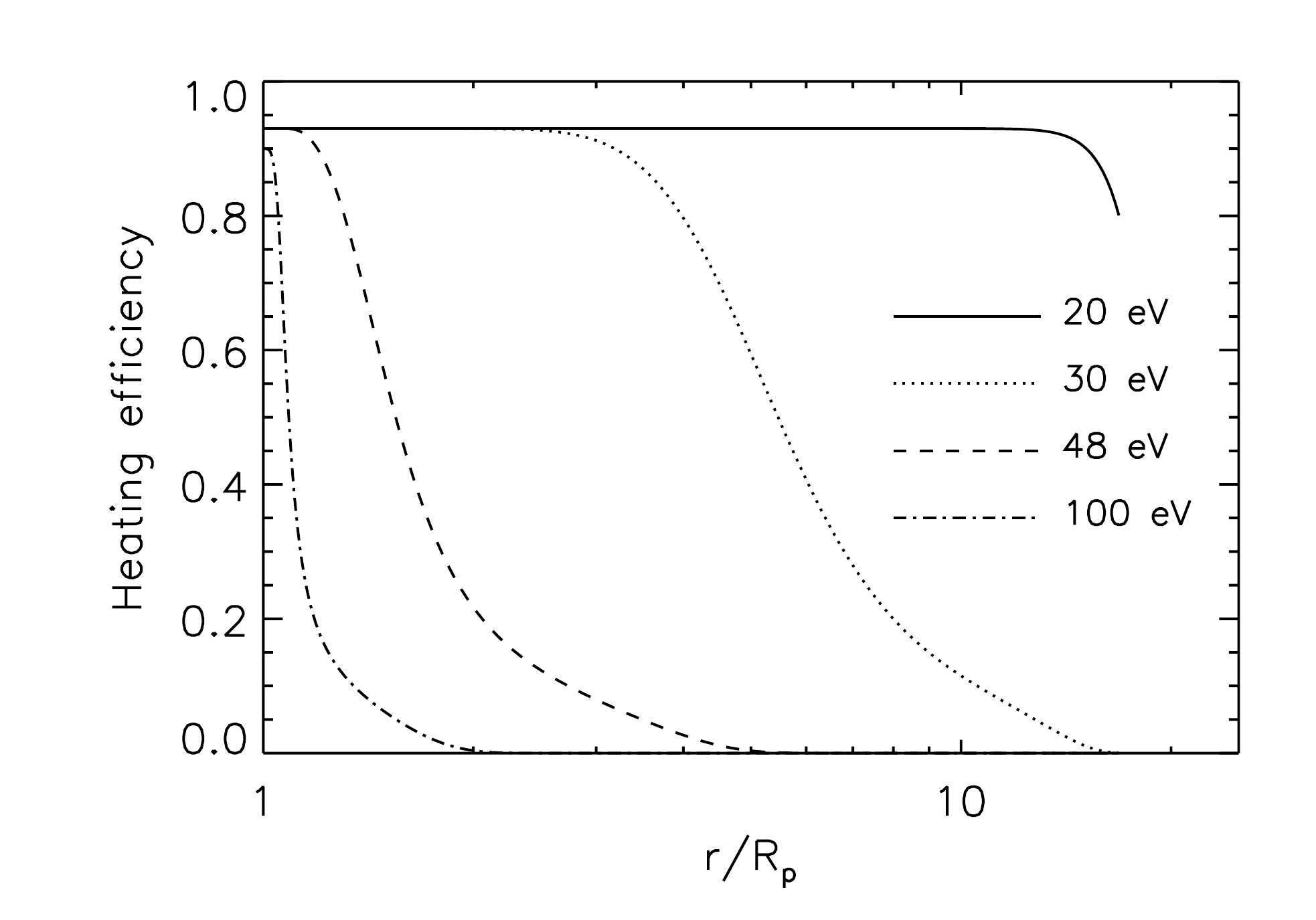}
  \caption{Heating efficiencies for photons of different energies (see text).}
  \label{fig:peffs}
\end{figure}

We calculated $\Lambda_{pe}$ for different photoelectrons based on the C1 model, and compared the result with the vertical length scale $H$ of the atmosphere.  The latter is either the scale height or $R_p$, depending on which is shorter.  When $\Lambda_{pe}/H \gtrsim$~0.005--0.01 we assumed that the heating efficiency decreases with altitude according to the transmission function for 50 eV photons.  The limiting value was chosen to obtain a rough agreement with the results of \citet{cecchi09} for 50 eV photons, and it implies that the heating efficiency approaches zero when $\Lambda_{pe}/H \gtrsim$~0.1.  We parameterized the result in terms of the column density of H based on the density profiles of the C1 model, and connected our results for low energy photoelectrons smoothly with the results of \citet{cecchi09} for photons with $E \gtrsim$~50 eV.  We then generated the C2 model from the C1 model with the new heating efficiencies.  Figure~\ref{fig:peffs} shows the resulting heating efficiencies for 20, 30, 48, and 100 eV photons.

\begin{figure}
  \centering
  \includegraphics[width=0.7\textwidth]{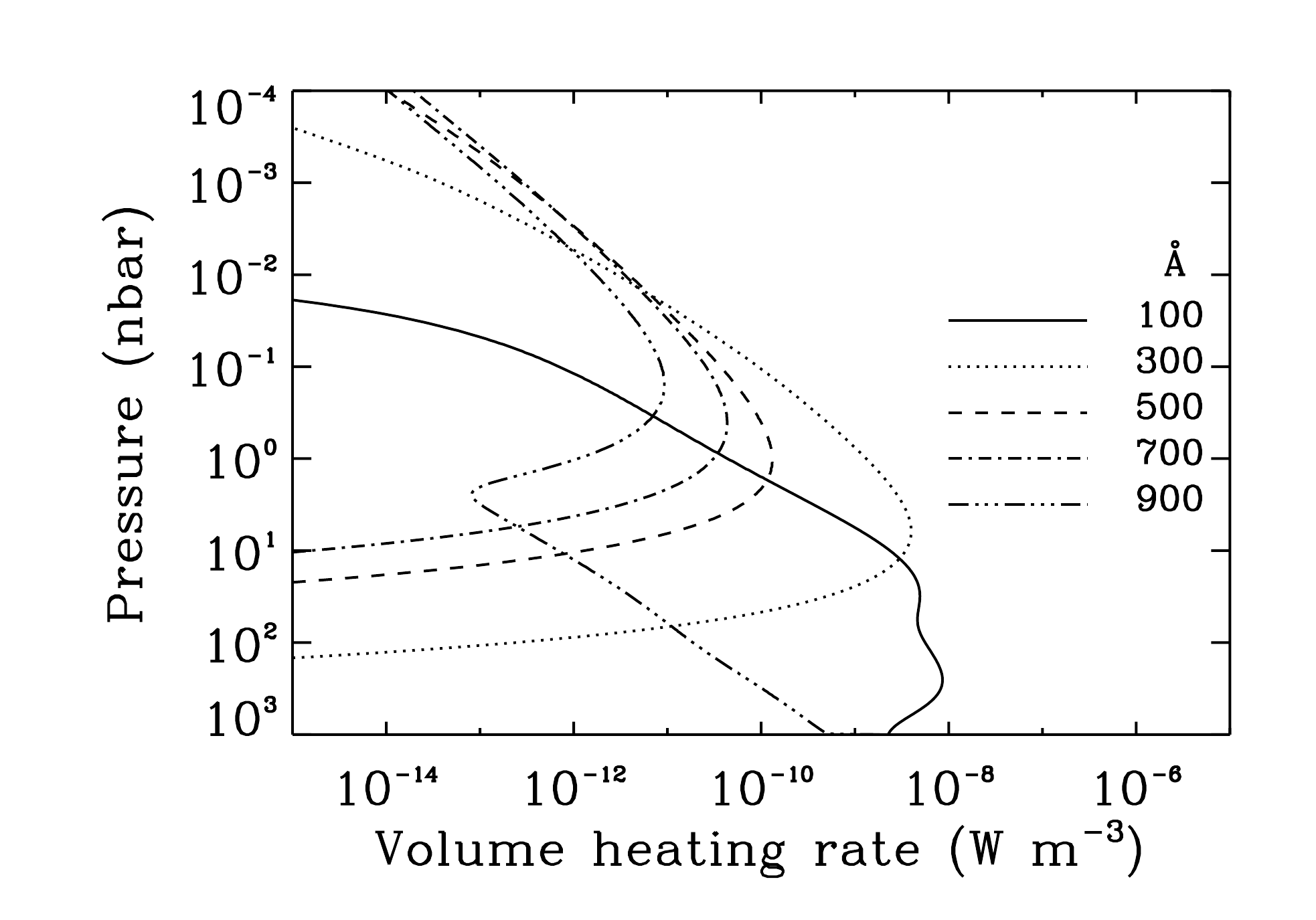}
  \caption{Volume heating rate as a function of pressure in the C2 model due to the absorption of stellar XUV radiation between 1 and 1000~\AA~in 200~\AA~bins.}
  \label{fig:penetration}
\end{figure}

Figure~\ref{fig:penetration} shows the volume heating rate due to EUV photons of different energies as a function of pressure based on the C2 model.  The maximum temperature of 12,000 K is reached near 1.5 R$_p$ ($p =$~0.6 nbar).  This region is heated mainly by EUV photons with wavelengths between 200 and 900~\AA~($E =$~14--62 eV).  The saturation heating efficiency of 93 \% for these photons is higher than the corresponding heating efficiency in the Jovian thermosphere \citep{waite83}.  This is because of strong ionization that leads to frequent Coulomb collisions between photoelectrons and thermal electrons.  Radiation with wavelengths shorter than 300~\AA~($E >$~40 eV) or longer than 912~\AA~(13.6 eV) penetrates past the temperature peak to the lower atmosphere.  The heating efficiency for photons with $E >$~25 eV approaches zero at high altitudes where heating is mostly due to low energy EUV photons.  The net heating efficiency for the C2 model is $\eta_{\text{net}} =$~0.44 (Table~\ref{table:models}), which is close to the H50 model.  The location of the peak and maximum temperature in the C2 model agree with the H50 model, but the temperature at higher altitudes in the C2 model decreases much more rapidly with altitude than in the H50 model.   

\begin{figure}
  \centering
  \includegraphics[width=0.7\textwidth]{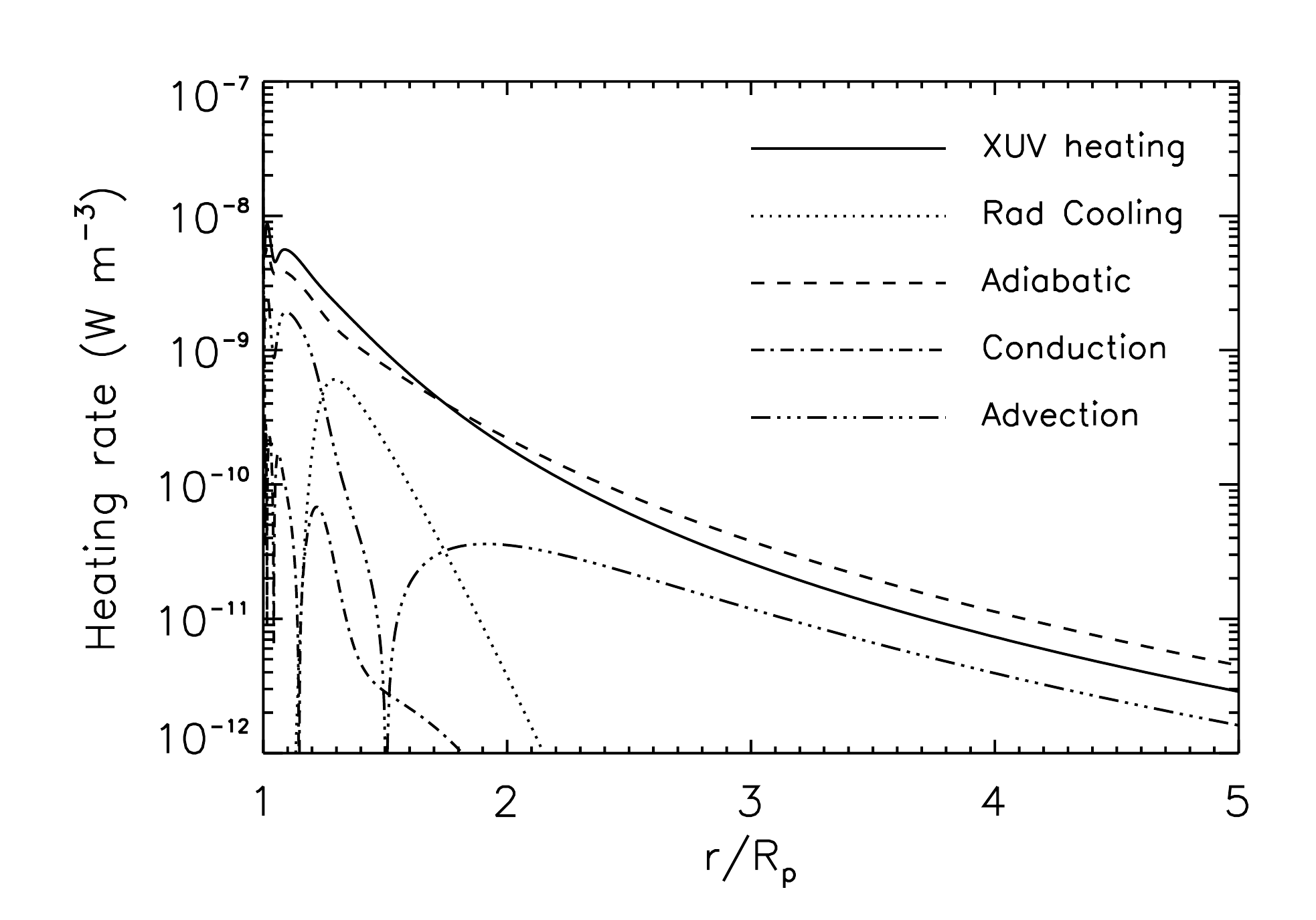}
  \caption{Volume heating rate based on the C2 model (absolute values).  Advection acts as a cooling mechanism below 1.5 $R_p$ and a heating mechanism above this level.}
  \label{fig:c2energy}
\end{figure}

Figure~\ref{fig:c2energy} shows the terms in the energy equation based on the C2 model.  In line with previous studies, stellar heating is mainly balanced by adiabatic cooling.  Advection cools the atmosphere at low altitudes below the temperature peak, whereas at higher altitudes it acts as a heating mechanism.  In fact, above 2 $R_p$ the adiabatic cooling rate is higher than the stellar heating rate because thermal energy is transported to high altitudes by advection from the temperature peak.  The radiative cooling term that is centered near 1.3 $R_p$ arises from recombination following thermal ionization.  Recombination following photoionization is included implicitly in the model and the rate is not included in the output.  Conduction is not significant at any altitude in the model.  We note that the rates displayed in Figure~\ref{fig:c2energy} balance to high accuracy, thus implying that the simulation has reached steady state.      

The differences between the C1 and C2 models are subtle.  The peak temperatures are similar, and the temperature profiles generally coincide below 3 $R_p$.  Above this radius the temperature in the C2 model decreases more rapidly with altitude than in the C1 model and subsequently the sonic point moves to higher altitudes above the model domain.  The results indicate that the assumption of a constant photoelectron heating efficiency is appropriate below 3 $R_p$ whereas at higher altitudes it changes the nature of the solution.  This should not be confused with the assumption of a constant $\eta_{\text{net}}$, which leads to a different temperature profile when compared with either C1 and C2 (see Figure~\ref{fig:gentempvel}).  In general, the maximum and mean temperatures in models C1--C4 are relatively similar.  Thus we conclude that the mean temperature in the thermosphere of HD209458b is approximately 7,000 K and the maximum temperature is 10,000--12,000 K.   
            
The substellar tide is included in the C3 model.  We included it mainly to compare our results with previous models \citep{garciamunoz07,penz08,murrayclay09}.  The substellar tide is not a particularly good representation of the stellar tide in a globally averaged sense.  In reality, including tides in the models is much more complicated than simply considering the substellar tide \citep[e.g.,][]{trammell11}.  Compared to the C1 model, the maximum temperature in the C3 model is cooler by $\sim$1,000 K and at high altitudes the C3 model is cooler by 1,000--2,000 K.  The velocity is faster and hence adiabatic cooling is also more efficient.  The substellar tide drives supersonic escape \citep[see also,][]{penz08} and the sonic point in the C3 model is at a much lower altitude than in the C1 model (see Section~\ref{subsc:sonicpoint}).  However, it is not clear how the sonic point behaves as a function of latitude and longitude.  Given that the tide is also likely to induce horizontal flows, it cannot be included accurately in 1D models.                    

\citet{murrayclay09} argued that radiative cooling due to the emission of Lyman~$\alpha$ photons by excited H is important on close-in EGPs.  The photons are emitted when the 2p level of H, which is populated by collisions with thermal electrons and other species, decays radiatively.  We included this cooling mechanism in the C4 model by using the rate coefficient from \citet{glover07} that includes a temperature-dependent correction to the rate coefficient given by \citet{black81}.  We also included an additional correction factor of 0.1 based on detailed level population and radiative transfer calculations by \citet{menager11}.  The effect of Lyman $\alpha$ cooling is largest near the temperature peak where the C4 model is 1,500 K cooler than the C1 model, but generally the difference is not large.  We note that the H Lyman $\alpha$ cooling rate here cannot be generalized as such to other EGPs because the level populations and opacities depend on the thermal structure and composition of the atmosphere.     

\subsubsection{Critical points}
\label{subsc:sonicpoint}   

As we have pointed out, the location of the sonic point depends on the energy equation through the temperature profile.  Here we show that the use of the isothermal approximation in estimating the location of the sonic point can lead to significant errors unless the atmosphere really is isothermal.  The inviscid continuity and momentum equations can be combined to give an expression for the critical point $\xi_c$ of a steady-state solution \citep{parker65}:
\begin{equation}
-\frac{\textrm{d}}{\textrm{d} \xi} \left( \frac{c^2}{\xi^2} \right) = -\frac{1}{\xi^2} \frac{\textrm{d} c^2}{\textrm{d} \xi} + \frac{2 c^2}{\xi^3} = \frac{W^2}{\xi^4} 
\label{eqn:parkerpoint}
\end{equation}
where $\xi = r/r_0$, $c = \sqrt{k T/m}$ is the isothermal speed of sound, $W = GM_p/r_0$, and $m$ is the mean atomic weight.  It is often assumed that the vertical velocity at the critical point is given by $v^2 = c^2 (\xi_c)$ so that the critical point coincides with the isothermal sonic point \citep{parker58}.  However, \citet{parker65} suggested that subsonic solutions are also possible if the density at the base of the flow exceeds a critical value determined from the energy equation.  In fact, he argued that conduction at the base of the corona may not be sufficient to drive a supersonic solar wind.  This led him to suggest that supersonic expansion is possible only if there is significant heating of the corona over large distances above the base.  

For an isothermal atmosphere with a temperature $T_0$, equation (\ref{eqn:parkerpoint}) reduces to the famous result for the altitude of the sonic point \citep{parker58}:
\begin{equation}
\xi_c = \frac{W^2}{2 c_0^2}    
\label{eqn:parkerpoint2}   
\end{equation}  
where $W^2/c_0^2$ is the thermal escape parameter $X_0$ at the lower boundary $r_0$.  The isothermal sonic point in the C1 model is located at 7.2 R$_p$ where $c(\xi_c) =$~7.2 km~s$^{-1}$.  The volume averaged temperature of the C1 model below this point is approximately 7,100 K.  Assuming that $r_0 =$~$R_p$, $T_0 =$~7,100 K, and $m = m_{\textrm{H}}$, $X_0 \sim$~16 and equation~(\ref{eqn:parkerpoint2}) yields $\xi_c \sim$~8.  In this case the analytic result agrees fairly well with the hydrodynamic model if one accounts for partial ionization of the atmosphere by assuming that the mean atomic weight\footnote{The mean atomic weight can be less than 1 because electrons contribute to the number density but not significantly to the mass density.} is $m =$~0.9~$m_H$.  

On the other hand, the isothermal sonic point in the C2 model is at 15.4 $R_p$ where $c(\xi_c) =$~4.1 km~s$^{-1}$.  This is because the temperature gradient of the model is steeper than the corresponding gradient in the C1 model.  The volume averaged mean temperature below 15 $R_p$ in the C2 model is 3,900 K.  With this temperature and $m = m_H$, equation~(\ref{eqn:parkerpoint2}) predicts a sonic point at 14.6 $R_p$.  However, at 15 $R_p$ the atmosphere is mostly ionized and $m =$~0.6~$m_H$.  With this value, the sonic point from equation~(\ref{eqn:parkerpoint2}) would be located at 8.8 $R_p$.  These examples show that there are significant caveats to using equation~(\ref{eqn:parkerpoint2}) to estimate the altitude of the sonic point on close-in EGPs without accurate knowledge of the temperature and density profiles.  A variety of outcomes are possible and it is difficult to develop a consistent criteria for choosing values of $T$ and $m$ that would produce satisfactory results.   

Another problem is that the atmosphere is not isothermal.  In fact, the temperature gradient above the heating peak in models C1--C4 (Table~\ref{table:models}) is relatively steep, and in some cases it approaches the static adiabatic gradient ($T \propto r^{-1}$) as defined by \citet{chamberlain61}.  Assuming that the temperature profile can be fitted with $c^2 = c_0^2/\xi^{\beta}$ above the heating peak, the estimated values of $\beta$ for the C1 and C2 models are 0.7 and 0.9, respectively.  We note that the velocity in the C1 model exceeds the isentropic speed of sound ($c_{\gamma} = \sqrt{\gamma k T/m}$ where $\gamma =$~5/3) at 9.8 $R_p$ where $c_{\gamma} =$~8.7 km~s$^{-1}$.  This altitude is significantly higher than the altitude of the isothermal sonic point.  The velocity in the C2 model does not exceed the isentropic speed of sound below the upper boundary of 16 $R_p$.  Thus the temperature profile has a significant impact on the nature of the solution and the escape mechanism.  This means that estimating the altitude of the sonic point without observations and detailed models for guidance is almost certain to produce misleading results. 

Past models for the upper atmosphere of HD209458b have predicted a variety of altitudes for the sonic point.  On the other hand, \citet{yelle04} pointed out that stellar heating in the thermosphere is mostly balanced by adiabatic cooling and our calculations confirm this.  \citet{parker65} argued that the critical point stretches to infinity when $\beta \rightarrow$~1 i.e., as the temperature gradient is close to adiabatic.  Based on this, we should perhaps expect that the sonic point on close-in EGPs is located at a fairly high altitude.  This is confirmed by our hydrodynamic simulations.  In all of our models except for one, the sonic point is located significantly above 5 $R_p$.  The exception is the C3 model, which includes the substellar tide.  The isentropic sonic point in this model is located at 3.9 $R_p$ where $c_{\gamma} =$~8.2 km~s$^{-1}$.  This is because the substellar tide leads to a lower effective value of the potential $W$.  However, the tidal potential depends on latitude and longitude, and the substellar results cannot be generalized globally.       

\subsubsection{Mass loss rates}
\label{subsc:massloss}

Here we evaluate the mass loss rates based on our models.  We define the mass loss rate simply as:
\begin{equation}
\dot{M} = 4 \pi r^2 \rho v.
\end{equation}
We note that the solar spectrum that we used in this study contains the total flux of  4~$\times$~10$^{-3}$ W~m$^{-2}$ at wavelengths shorter then 912~\AA~(the ionization limit of H) when normalized to 1 AU.  This value is close to the average solar flux of 3.9~$\times$~10$^{-3}$ W~m$^{-2}$ at the same wavelengths \citep[e.g.,][]{ribas05}.  In order to simulate a global average, we divided the flux by a factor of 4 in the model.  This means that the incident flux on HD209458b at 0.047 AU with wavelengths shorter than 912 \AA~in our model is 0.45 W~m$^{-2}$.  The net heating efficiencies given in Table~\ref{table:models} are based on this value.  

Considering first the models with constant $\eta_{\text{net}}$ ranging from 0.1 to 1 (see Section~\ref{subsc:gendep}), the mass loss rate varies almost linearly with $\eta_{\text{net}}$ from 10$^7$ kg~s$^{-1}$ and 10$^8$ kg~s$^{-1}$ while the pressure averaged temperature below 3 $R_p$ changes only by 1,500 K.  This is because in a hydrodynamic model such as ours the net energy has nowhere else to go but adiabatic expansion and cooling, and thus escape is energy-limited.  The bulk of the energy is absorbed below 3 $R_p$, and the mass loss rate is largely set by radiative transfer in this region.  The mass loss rate for HD209458b predicted by the C2 model is 4.1~$\times$~10$^7$ kg~s$^{-1}$ ($\eta_{\text{net}} =$~0.44).  The C3 model has the highest mass loss rate, although this rate is only higher by a factor of 1.13 than the mass loss rate in the C1 model.  Thus we predict a mass loss rate of 4--6~$\times$~10$^{7}$ kg~s$^{-1}$ from HD209458b based on the average solar flux at 0.047 AU.    

\citet{garciamunoz07} demonstrated that the mass loss rate is insensitive to the upper boundary conditions even when they have a large impact on the temperature and velocity profiles, particularly at high altitudes.  Indeed, complex hydrodynamic models are not required to calculate mass loss rates under energy-limited escape as long as reasonable estimates of the net heating efficiency are available.  It is also important to note that the current estimates of mass loss rates based on the observations \citep[e.g.,][]{vidalmadjar03} are not direct measurements.  Instead, they are all based on different models.  However, models that predict the same mass loss rate can predict different transit depths and models predicting different mass loss rates can match the observations equally well.  Thus the models should not be judged on how well they agree with published mass loss rates but rather on how well they agree with the observed density profiles or transit depths.  Hydrodynamic models with realistic heating rates are required to match the observations, and the mass loss rate then follows.   

The globally averaged mass loss rate of about 4--6~$\times$~10$^7$ kg~s$^{-1}$ from HD209458b agrees well with similar estimates calculated by \citet{yelle04,yelle06} and \citet{garciamunoz07}, respectively, but it is significantly larger than the value calculated by \citet{murrayclay09}.  These authors report a mass loss rate of 3.3~$\times$~10$^7$ kg~s$^{-1}$ based on the substellar atmosphere.  When multiplied by 1/4 this corresponds to a global average rate of about 8.3~$\times$~10$^6$ kg~s$^{-1}$.  However, the substellar mass loss rate is also enhanced by tides, so a comparable global average taking this into account would be even less than 8.3~$\times$~10$^6$ kg~s$^{-1}$, which is already roughly a factor of 6 smaller than our calculations.

The \citet{murrayclay09} models differ in many respects from the models described here including the treatment of boundary conditions and radiative cooling, the numerical approach, and the adoption of a gray approximation for stellar energy deposition.  In order to explore the reason for the disagreement in escape rates, we have modified our model to implement the gray assumption by using the approach described in \citet{murrayclay09} (see Section~\ref{subsc:denprofs}).  Specifically, we adopted an incident flux\footnote{By chance the incident flux is equal to the mean solar flux divided by 4 that we used as a `globally averaged' value.  Here, however, it is taken to be the substellar value.} of 0.45 W~m$^{-2}$ and a mean photon energy of 20 eV.  The mass loss rate based on the substellar atmosphere for this model is 2.8~$\times$~10$^7$ kg~s$^{-1}$, in good agreement with the \citet{murrayclay09} results.  Thus, the difference between the \citet{murrayclay09} models and the others is due to the gray assumption, and the fact that they estimated the incident flux on HD209458b based on the solar flux integrated between 13 eV and 40 eV.  This energy range contains only about 25 \% of the total solar flux at energies higher than 13.6 eV. 

Although not discussed by \citet{murrayclay09}, the restricted energy range is likely an attempt to account for the fact that the absorption cross section decreases with energy implying that photons of sufficiently high energy will be absorbed too deep in the atmosphere to affect escape or the thermal structure, or composition of the thermosphere.  Whether this is true, however, depends on the composition and temperature of the atmosphere.  The gray assumption also fails to include the fact that the net heating efficiency increases with higher photon energy.  These difficulties highlight the basic problem with a gray model, that the results can only be accepted with confidence if verified by a more sophisticated calculation or direct observations.

\subsubsection{Constraints from kinetic theory}
\label{subsc:boundaries}

Hydrodynamic models should be consistent with kinetic theory of rarefied media even if the modeled region is below the exobase.  This is because the atmosphere is escaping to space, and the density decreases with altitude, falling below the fluid regime at some altitude above the exobase.  Therefore the conditions in the exosphere affect the flow solutions even below the exobase.  Inappropriate use of the hydrodynamic equations can lead one to overestimate the flow velocity and mass loss rate, and these errors also affect the temperature and density profiles.  Thus it is important to demonstrate that the hydrodynamic solutions agree with constraints from kinetic theory \citep[e.g.,][]{volkov11a,volkov11b}.   

As an example, we calculated $Kn_0$ and $X_0$ (see Section~\ref{subsc:upbound}) based on the C1 and C2 models.  The Knudsen number $Kn$ depends on the mean collision frequency, and it is much smaller than unity at all altitudes below 16 $R_p$.  Thus the exobase is located above the model domain \citep[see also][]{murrayclay09}.  Calculating values for $X_0$ is complicated by the broad stellar heating profile.  We consider the region where stellar heating is negligible to be where the flux of energy  
\begin{equation}
E_{\infty} = F_c \left( c_p T + \frac{1}{2} v^2 - \frac{G M_p}{r} \right) - \kappa r^2 \frac{\partial T}{\partial r}
\end{equation}
is approximately constant.  This criteria is consistent with the equations of motion, and it means that $r_0$ that should be used to calculate $X_0$ is above the upper boundary of our model because significant stellar heating persists at all altitudes.  Thus we evaluated values of $X$ near the upper boundary for guidance.  We also calculated the values with both the mass of the proton ($X_{\textrm{H}}$) and the mean atomic weight ($X$).

In the C1 model, $X_{\textrm{H}}$ decreases with altitude, and above 11.4 $R_p$ it has values of less than 3.  The mean atomic weight near the upper boundary is $\sim$0.6 amu, and thus the general value of $X <$~2 above 11.1 $R_p$.  The sonic point in the C1 model is below 11 $R_p$, and it is in a region where stellar heating is significant.  In the C2 model, both $X$ and $X_{\text{H}}$ are greater than 3 at all altitudes below 16 $R_p$.  In fact, $X$ increases with altitude above 9 $R_p$ because the temperature gradient parameter exceeds unity.  Thus the values $X$ in the C1 and C2 models are consistent with the difference in altitude between the sonic points in these models (see Section~\ref{subsc:sonicpoint}).  Indeed, our results show, in line with Parker's original ideas about the solar wind, that supersonic escape is possible if there is significant heating of the atmosphere over large distances above the temperature peak.  Such heating flattens the temperature gradient and brings the sonic point closer to the planet.   

We note that there are some caveats to applying the simple criteria based on $Kn_0$ and $X_0$ to close-in extrasolar planets.  The upper atmospheres of these planets are strongly ionized, and the DSMC simulations of \citet{volkov11a,volkov11b} apply only to neutral atmospheres.  Partly due to ionization, the collision frequencies in the thermospheres of close-in planets are also high.  Further, the atmospheres are affected by a broad stellar heating profile in altitude whereas the DSMC calculations do not include any diabatic heating.  However, the results of \citet{volkov11a,volkov11b} also indicate that consistency with kinetic theory can be enforced approximately by applying the modified Jeans conditions to the hydrodynamic model at some altitude close to the exobase.  This result is likely to be more general, and it applies to ionized atmospheres as long as ambipolar diffusion is taken into account (see Section~\ref{subsc:upbound}).    

\begin{figure}
  \centering
  \includegraphics[width=0.7\textwidth]{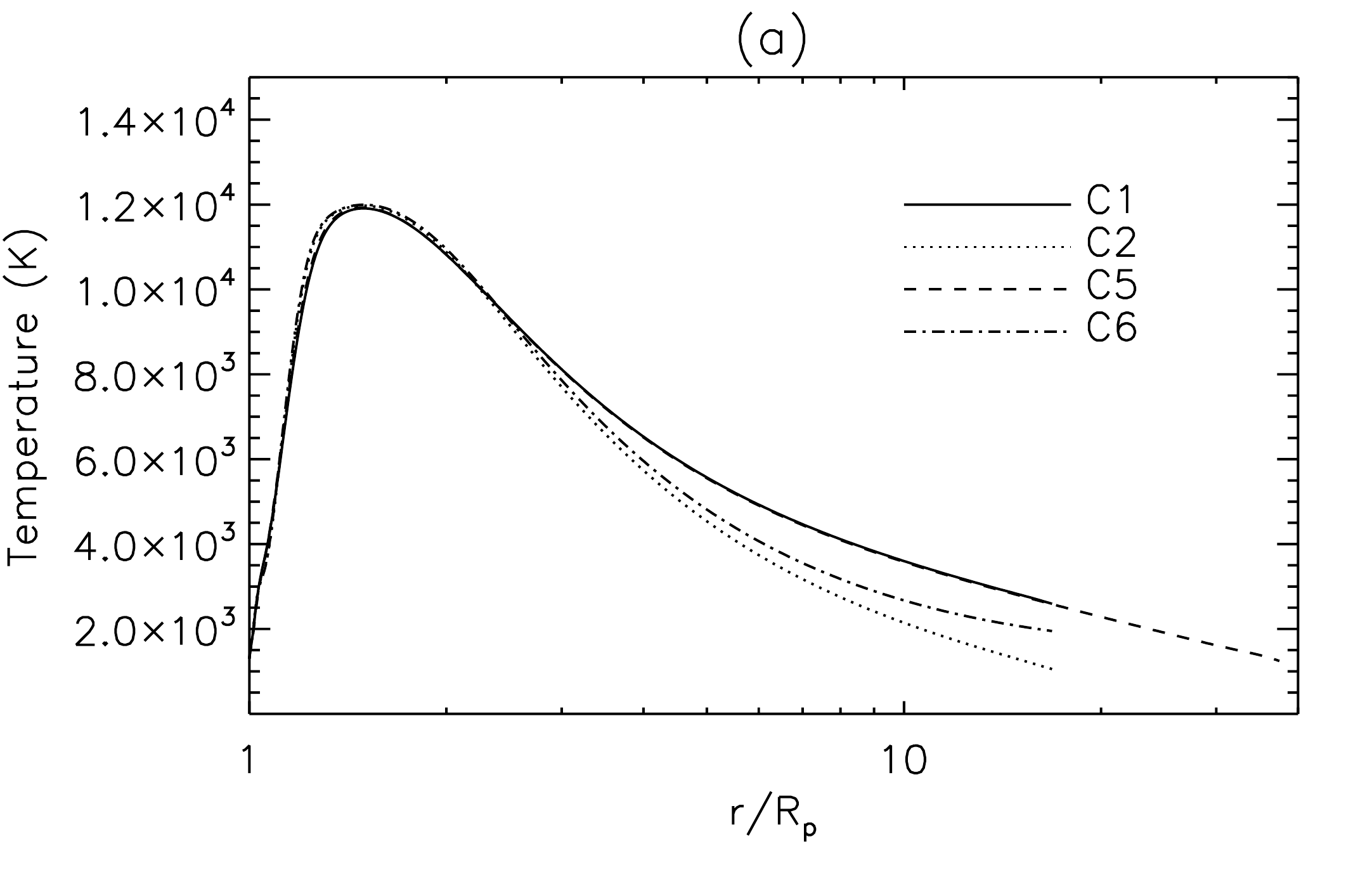}
  \includegraphics[width=0.7\textwidth]{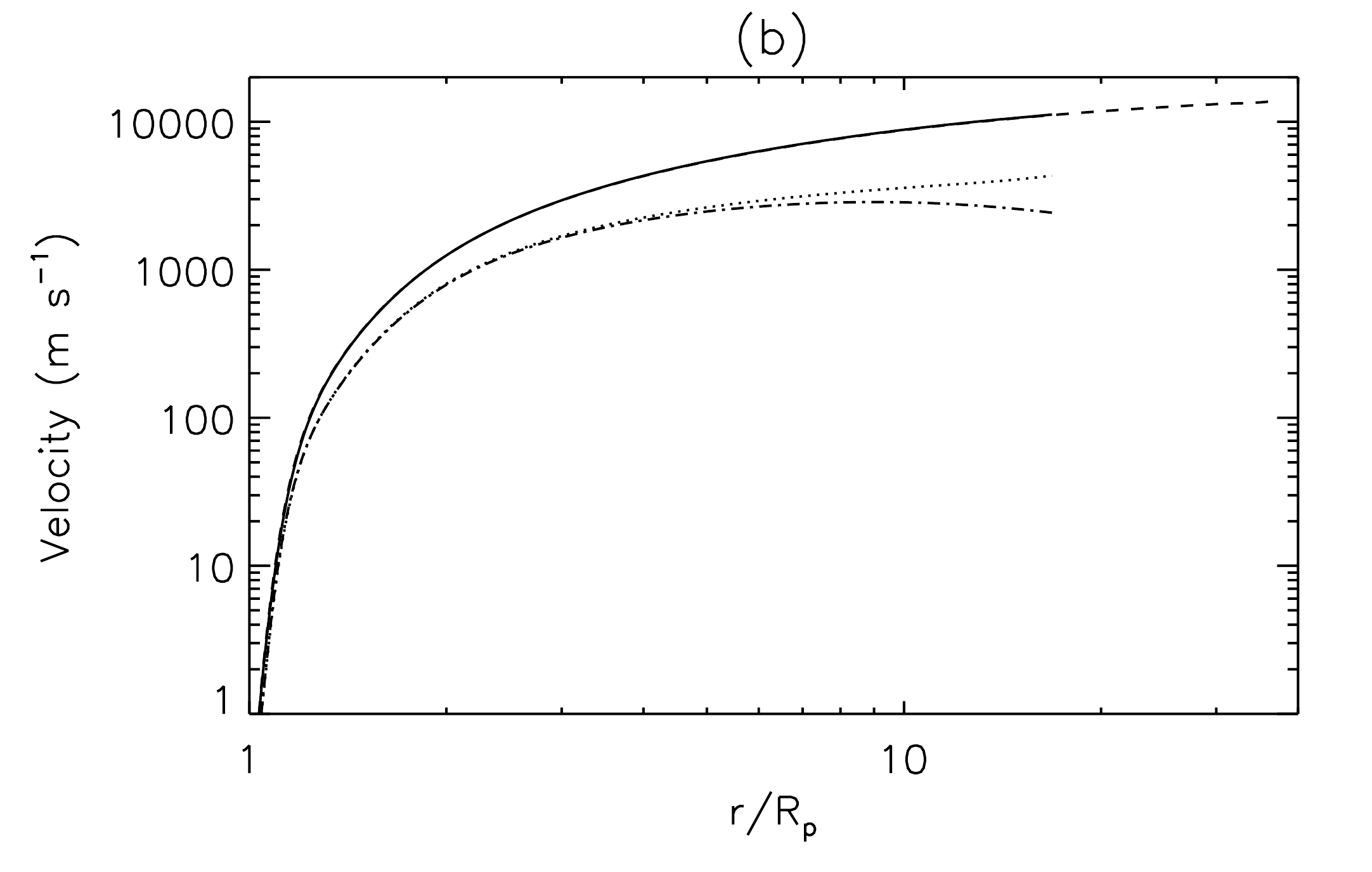}
  \caption{Temperature (a) and velocity (b) profiles in the upper atmosphere of HD209458b based on models with extrapolated and modified Jeans upper boundary conditions (see Table~\ref{table:models} for the input parameters).}
  \label{fig:modjeans}
\end{figure}

We compared the temperature and velocity profiles from the C1 and C2 models with results from similar models C5 and C6 that use the modified Jeans conditions.  Note again that our version of the modified Jeans conditions includes the polarization electrostatic field that is required in strongly ionized media.  Figure~\ref{fig:modjeans} shows the temperature and velocity profiles from the models.  There is no difference between the C5 model and the C1 model as long as the upper boundary of the C5 model is at a sufficiently high altitude.  In this case we extended it to 36 $R_p$.  When the upper boundary is placed at lower altitudes, the flow decelerates and the temperature increases near the upper boundary.  A comparison between the C2 and C6 models provides an example of the difference that can arise when the modified Jeans boundary conditions are used significantly below the exobase.  A better agreement is achieved if the upper boundary of the C6 model is placed at a slightly higher altitude.  In summary, we have shown that the C1 and C2 models are both consistent with kinetic theory.   

We note that extending the models to 16 $R_p$ or higher is not necessarily justified because it ignores the complications arising from the possible influence of the stellar tide, the stellar wind, and interactions of the flow with the magnetic field of the planet or the star.  We placed the upper boundary at a relatively high altitude to make sure that the boundary is well above the region of interest, but generally we do not consider our results to be accurate above 3--5 $R_p$.  Instead, our results provide robust lower boundary conditions for multidimensional models of the escaping material outside the Roche lobe of the planet.  Such models often cannot include detailed photochemical or thermal structure calculations.  The results from the more complex models can then be used to constrain the upper boundary conditions in 1D models.  This type of an iteration is a complex undertaking, and it will be pursued in future work.   

\subsection{Density profiles}
\label{subsc:denprofs}

In this section we provide a qualitative understanding of the density profiles and transition altitudes that affect the interpretation of the observations.  Based on the gas giants in the solar system it might be expected that heavy species undergo diffusive separation in the thermosphere.  If this were the case, the transit depths in the O I, C II, and Si III lines \citep{vidalmadjar04,linsky10} should not be significantly higher than the transit depth at visible wavelengths.  It is therefore important to explain why diffusive separation does not take place in the thermosphere of HD209458b, and to clarify why H and O remain mostly neutral while C and Si are mostly ionized.  Also, doubly ionized species such as Si$^{2+}$ are not common in planetary ionospheres, and their presence needs to be explained.  In order to do this, we modeled the ionization and photochemistry of the relevant species, and prove that diffusive separation does not take place.

\begin{figure}
  \centering
  \includegraphics[width=0.7\textwidth]{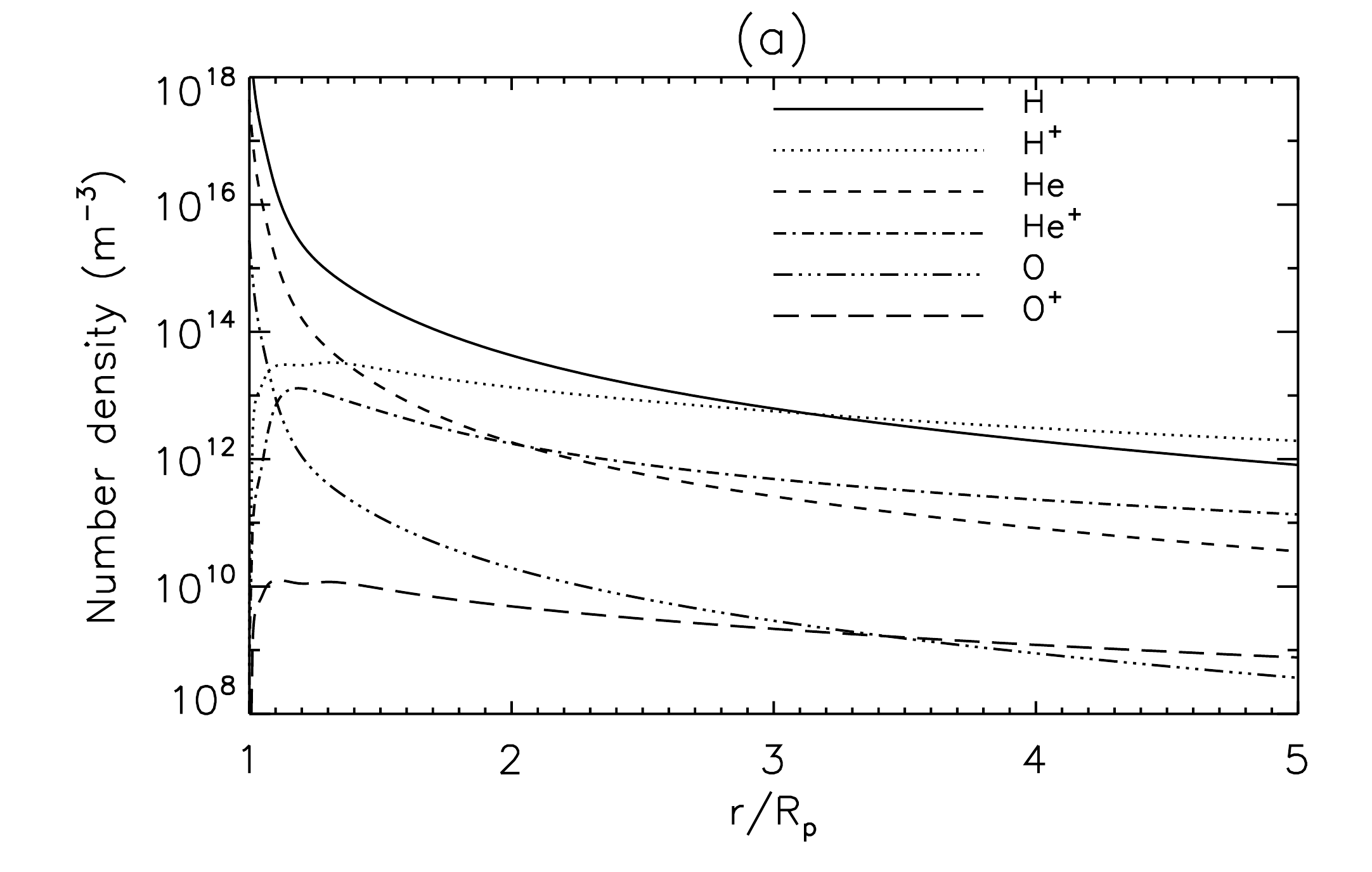}
  \includegraphics[width=0.7\textwidth]{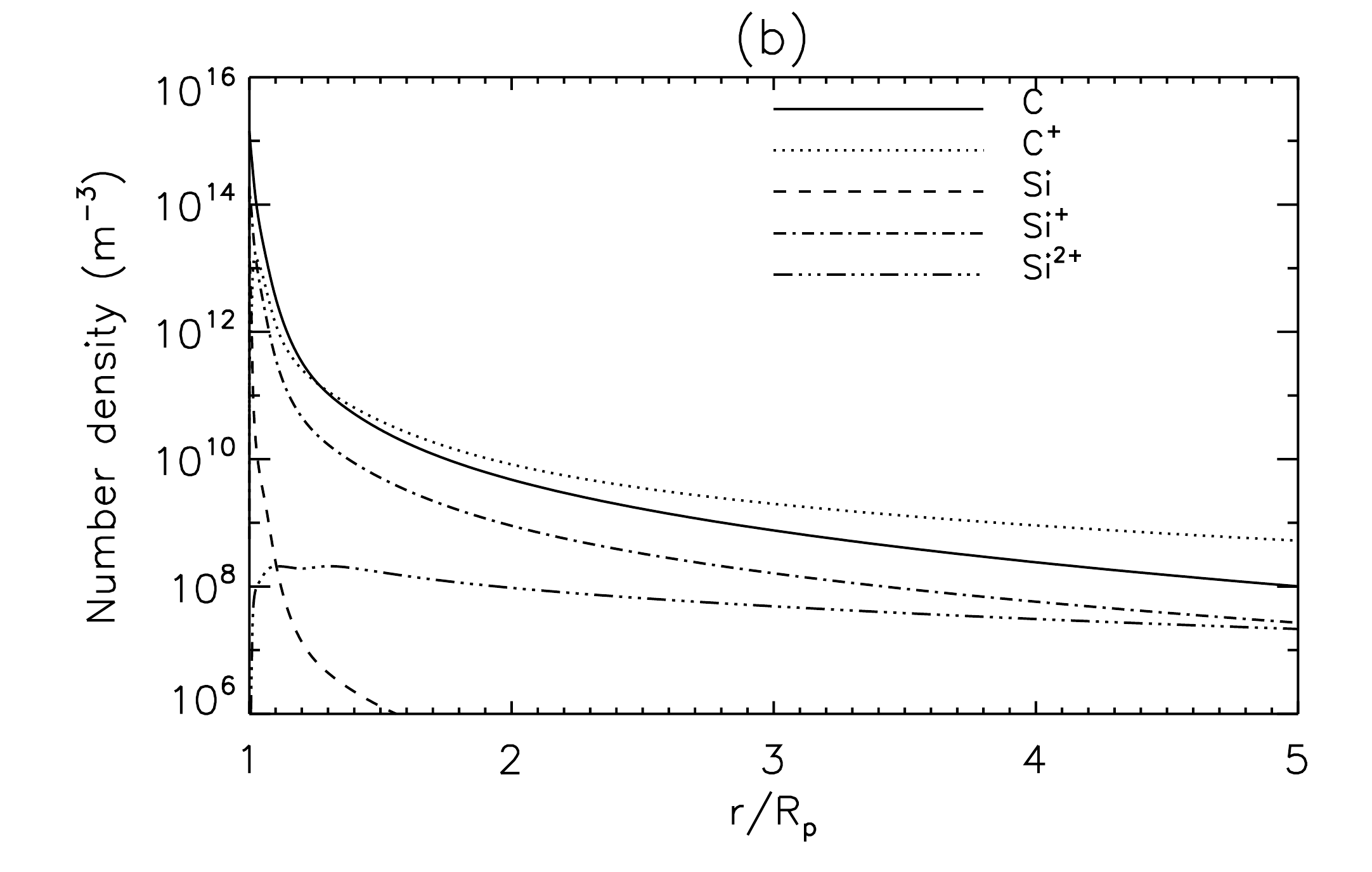}
  \caption{Density profiles in the upper atmosphere of HD209458b based on the C2 model (see Table~\ref{table:models} for the input parameters).}
  \label{fig:denprofs}
\end{figure}

In order to illustrate the results, Figure~\ref{fig:denprofs} shows the density profiles of H, H$^+$, He, He$^+$, O, O$^+$, C, C$^+$, Si, Si$^+$, and Si$^{2+}$ from the C2 model.  The location of the H/H$^+$ transition obviously depends on photochemistry, but it also depends on the dynamics of escape.  With a fixed pressure at the lower boundary, a faster velocity leads to a transition at a higher altitude.  Thus the transition occurs near 3.1 $R_p$ in the C2 model whereas in the C1 and C3 models it occurs at 3.8 $R_p$ and 5 $R_p$, respectively.  These results disagree with \citet{yelle04} and \citet{murrayclay09} who predicted a lower transition altitude, but they agree qualitatively with the solar composition model of \citet{garciamunoz07}.  They also agree with the empirical constraints derived by \citet{koskinen10} from the observations.

\begin{figure}
  \centering
  \includegraphics[width=0.7\textwidth]{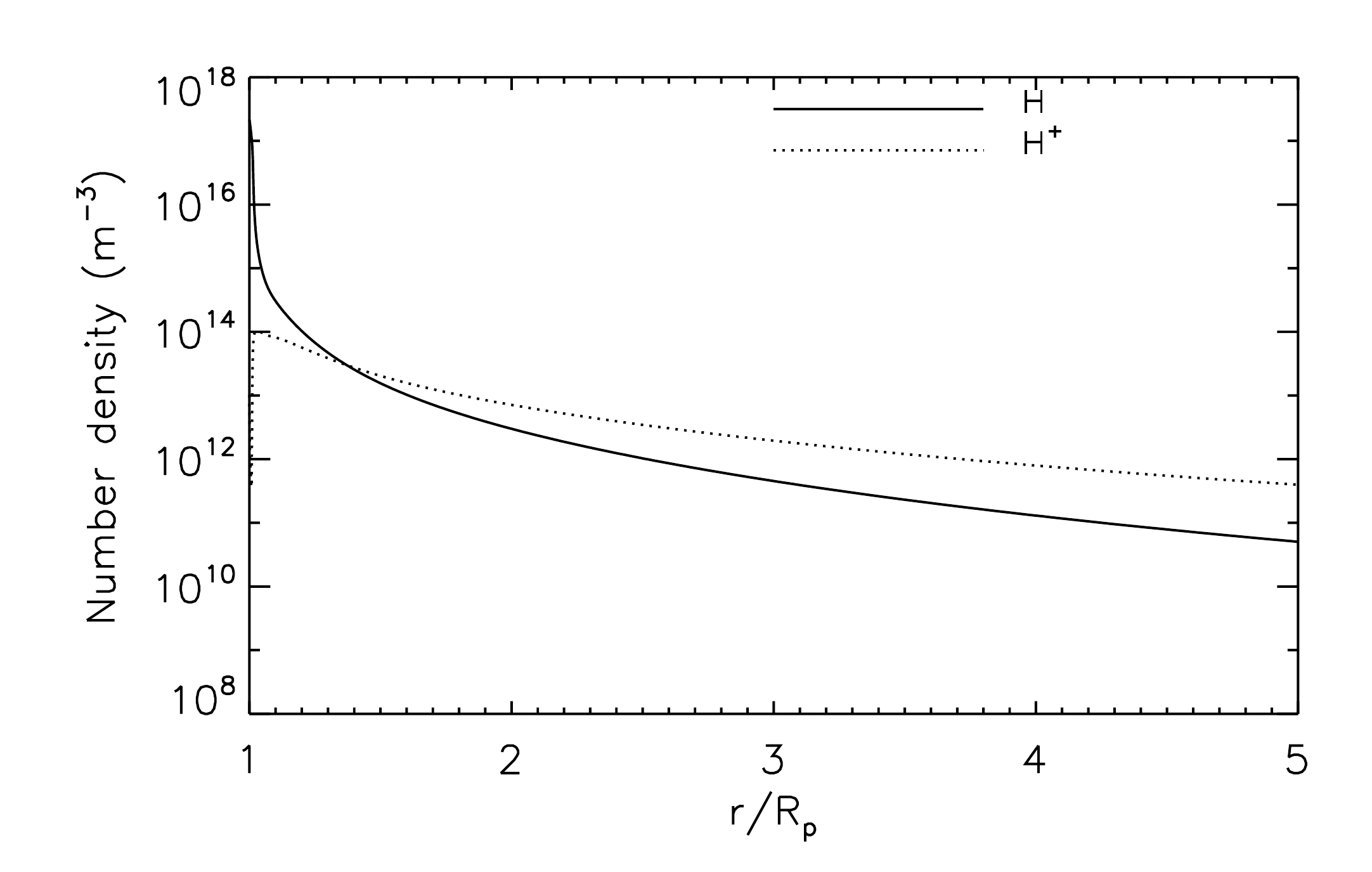}
  \caption{Density profiles of H and H$^+$ based on the MC09 model that is similar to that of \citet{murrayclay09} (see text).  Compared with our models (see Figure~\ref{fig:denprofs}), the H/H$^+$ transition occurs at a significantly lower altitude.  The difference arises from the lower boundary conditions and gray approximation to heating and ionization in the MC09 model.}  
  \label{fig:mc09}
\end{figure}

Once again, the differences between the earlier models and our work arise from different boundary conditions, and assumptions regarding heating rates and photochemistry.  We demonstrate this by reproducing the results of \citet{murrayclay09} with our model.  In order to do so, we set the lower boundary to 30 nbar with a temperature of 1,000 K, and included the substellar tide in the equations of motion.  We only included H, H$^+$, and electrons in the model, and used the recombination rate coefficient and Lyman~$\alpha$ cooling rate from \citet{murrayclay09}.  We also calculated the heating and ionization rates with the gray approximation by assuming a single photon energy of 20 eV for the stellar flux of 0.45 W m$^{-2}$ at the orbital position of HD209458b.  Figure~\ref{fig:mc09} shows the density profiles of H and H$^+$ based on this model (hereafter, the MC09 model).

The H/H$^+$ transition in the MC09 model occurs near 1.4 $R_p$.  If we replace the gray approximation with the full solar spectrum in this model, the H/H$^+$ transition moves higher to 2--3 $R_p$.  This is because photons with different energies penetrate to different depths in the atmosphere, extending the heating profile in altitude around the heating peak.  This is why the temperature at the 30 nbar level in the C2 model is 3,800 K and not 1,000 K.  In order to test the effect of higher temperatures in the lower thermosphere, we extended the MC09 model to $p_0 =$~1 $\mu$bar (with $T_0 =$~1,300 K) and again used the full solar spectrum for heating and ionization.  With these conditions, the H/H$^+$ transition moves up to 3.4 $R_p$, in agreement with the C2 model.  We conclude that the unrealistic boundary conditions and the gray approximation adopted by \citet{murrayclay09} and \citet{guo11} lead to an underestimated overall density of H and an overestimated ion fraction.  Thus their density profiles yield a H Lyman $\alpha$ transit depth of the order of 2--3 \% i.e., not significantly higher than the visible transit depth. 

We note that \citet{yelle04} also predicted a relatively low altitude of 1.7 $R_p$ for the H/H$^+$ transition -- despite the fact that his model does not rely on the gray approximation and the lower boundary is in the deep atmosphere.  The reason for the low altitude of the H/H$^+$ transition in this case is the neglect of heavy elements.  In the absence of heavy elements, H$_3^+$ forms near the base of the model and subsequent infrared cooling balances the EUV heating rates.  This prevents the dissociation of H$_2$ below the 10 nbar level.  In reality, reactions with OH and thermal decomposition dissociate H$_2$ near the 1 $\mu$bar level (see Section~\ref{subsc:photochemistry}) and cooling by H$_{3}^+$ is negligible at all altitudes.  It should be noted that even if H$_2$ does not initially dissociate, H$_3^+$ can be removed from the lower thermosphere in reactions with carbon and oxygen species \citep[e.g.,][]{garciamunoz07} unless these species undergo diffusive separation.  The subsequent lack of radiative cooling will then dissociate H$_2$ again near the 1 $\mu$bar level.  

In our models, charge exchange with oxygen (reactions R14 and R15 in Table~\ref{table:reactions}) dominates the photochemistry of H below 3 $R_p$ and charge exchange with silicon (R25, R26) is also important below 1.4 $R_p$.  These reactions are secondary in a sense that they require the ions to be produced by some other mechanism.  In fact, H$^+$ is mainly produced by photoionization (P1), although thermal ionization (R3) is also important near the temperature peak.  The production rates are mainly balanced by loss to radiative recombination (R1).  The net chemical loss timescale for H is longer than the timescale for advection above 1.7 $R_p$.  This allows advection from below to replenish H at higher altitudes.

The density profiles of O and O$^+$ are strongly coupled to H and H$^+$ by charge exchange \citep[see also][]{garciamunoz07}.  As a result, the O/O$^+$ transition occurs generally near the H/H$^+$ transition.  For instance, in the C2 model it is located near 3.4 $R_p$.  The same is not true of carbon.  The C/C$^+$ transition occurs at a much lower altitude than the H/H$^+$ and O/O$^+$ transitions.  For instance, in the C2 model it is located near 1.2 $R_p$.  C$^+$ is mainly produced by photoionization (P4), although thermal ionization (R8) and charge exchange with He$^+$ (R13) are also important near the temperature peak.  The production is balanced by loss to radiative recombination (R10).  The chemical loss timescale for C is shorter than the timescale for advection below 1.8 $R_p$.  Thus advection is unable to move the C/C$^+$ transition to altitudes higher than 1.2 $R_p$.

Silicon is almost fully ionized near the lower boundary of the model.  Much of the Si$^{+}$ below 4 $R_p$ is produced by charge exchange of Si with H$^+$, He$^+$, and C$^+$ (R22, R23, R24).  The low ionization potential of Si (8.2 eV) means that Si$^+$ can also be produced by thermal ionization (R18), and photoionization (P6) by stellar FUV radiation and X-rays that propagate past the EUV heating peak.  Above 4 $R_p$, Si$^+$ is mostly produced by photoionization.  \citet{linsky10} suggested that the balance of Si$^+$ and Si$^{2+}$ depends on charge exchange with H$^+$ and H, respectively, and our results confirm this.  However, the location of the Si$^+$/Si$^{2+}$ transition also depends on the dynamics.  For instance, in the C2 model it occurs near 5.8 $R_p$ while in the C1 model it occurs near 8.5 $R_p$.  Thus slow outflow and high temperatures favor Si$^{2+}$ as the dominant silicon species as long as the flux constant is high enough to overcome diffusive separation (Paper II).

We have now explained the presence of the atoms and ions that have been detected in the thermosphere of HD209458b.  Due to advection and charge exchange, H and O are predominantly neutral up to about 3 $R_p$ and give rise to the observed transit depths in the H Lyman $\alpha$ and O I lines.  Carbon, on the other hand, is ionized at a low altitude and thus C$^+$ is also detectable in the upper atmosphere.  Si$^+$ is the dominant silicon species below 5 $R_p$, but charge exchange with H ensures that there is also a significant abundance of Si$^{2+}$ in the atmosphere.  We note that these conclusions are only valid if the heavier species are carried along to high altitudes by the escaping hydrogen.  We show that this is the case below in Section~\ref{subsc:ionescape}.  

\subsubsection{The EUV ionization peak (EIP) layer}
\label{subsc:eip}

\citet{koskinen10b} explored the properties of the ionospheres of EGPs at different orbital distances from a Sun-like host star by using a hydrostatic general circulation model (GCM) that also includes realistic heating rates, photochemistry, and transport of constituents.  They predicted that the EIP layer on HD209458b is centered at 1.35 R$_p$ where the electron density is $n_e =$~6.4~$\times$~10$^{13}$ m$^{-3}$ and $x_e \sim$~3~$\times$~10$^{-2}$.  In the C2 model, the EIP layer is centered at 1.3 R$_p$ ($p =$~2 nbar) where $n_e =$~4.4~$\times$~10$^{13}$ m$^{-3}$ and $x_e =$~3.7~$\times$~10$^{-2}$.  The vertical outflow velocity at 1.3 R$_p$ is 90 m~s$^{-1}$.  Thus the results of \citet{koskinen10b} were not significantly affected by the simplifying assumptions of the GCM.  This means that hydrostatic GCMs can be extended to relatively low pressures as long as the escape rates are incorporated as boundary conditions.  

We also calculated the plasma frequency based on the electron densities in the C2 model.  This constrains the propagation of possible radio emissions from the ionosphere.  The ordinary plasma frequency $\omega_p / 2 \pi$ exceeds 12 MHz at all altitudes below 5 R$_p$ and reaches a maximum of almost 64 MHz in the EIP layer.  This presents a limitation on the detection of radio emissions from the ionospheres of close-in EGPs.  Any emissions that originate from the ionosphere at 1--5 R$_p$ and have frequencies lower than 10--70 MHz can be screened out by the ionosphere itself.  We note that a detection of radio emissions from an EGP has not yet been achieved \citep[e.g.,][]{bastian00,lazio07,lecavelier11,griessmeier11}.  Such a detection would be an important constraint on the magnetic field strength and the ionization state of the source region \citep[e.g.,][]{griesmeier07}.  Models of the ionosphere are required to predict radio emissions from the possible targets. 

\subsubsection{The escape of heavy atoms and ions}
\label{subsc:ionescape}

In this section we verify \textit{a posteriori} that the velocity and temperature differences between different species in the thermosphere of HD209458b are small.  This demonstrates that the single fluid approximation of the momentum and energy equations is valid, and that diffusive separation of the heavy species does not take place.  Our model includes velocity differences between different species by including all of the relevant collisions between them through the inclusion of diffusive fluxes in the continuity equations.  However, we have explicitly assumed that $T_n = T_i = T_e$, and this assumption in particular needs to be verified.  Also, the diffusion approach to the continuity equation is only valid if the velocity differences between the species are reasonably small. 

We calculated the collision frequencies based on the C2 model, and found that collisions with neutral H dominate the transport of heavy neutral atoms such as O below 3.5 $R_p$.  At altitudes higher than this, collisions with H$^+$ are more frequent.  In Paper II we demonstrate that a mass loss rate of 6~$\times$~10$^6$ kg~s$^{-1}$ is required to prevent diffusive separation of O (the heaviest neutral species detected so far) in the thermosphere.  The mass loss rate in our models is $\dot{M} >$~10$^7$ kg~s$^{-1}$ and thus O is dragged along to high altitudes by H.  On the other hand, collisions with H$^+$ dominate the transport of heavy ions such as Si$^+$ as long as the ratio [H$^+$]/[H]~$\gtrsim$~10$^{-4}$ (Paper II).  This explains why Coulomb collisions in our models are more frequent than heavy ion-H collisions at almost all altitudes apart from the immediate vicinity of the lower boundary.  These collisions are much more efficient in preventing diffusive separation than collisions with neutral H.  

\begin{figure}
  \centering
  \includegraphics[width=0.7\textwidth]{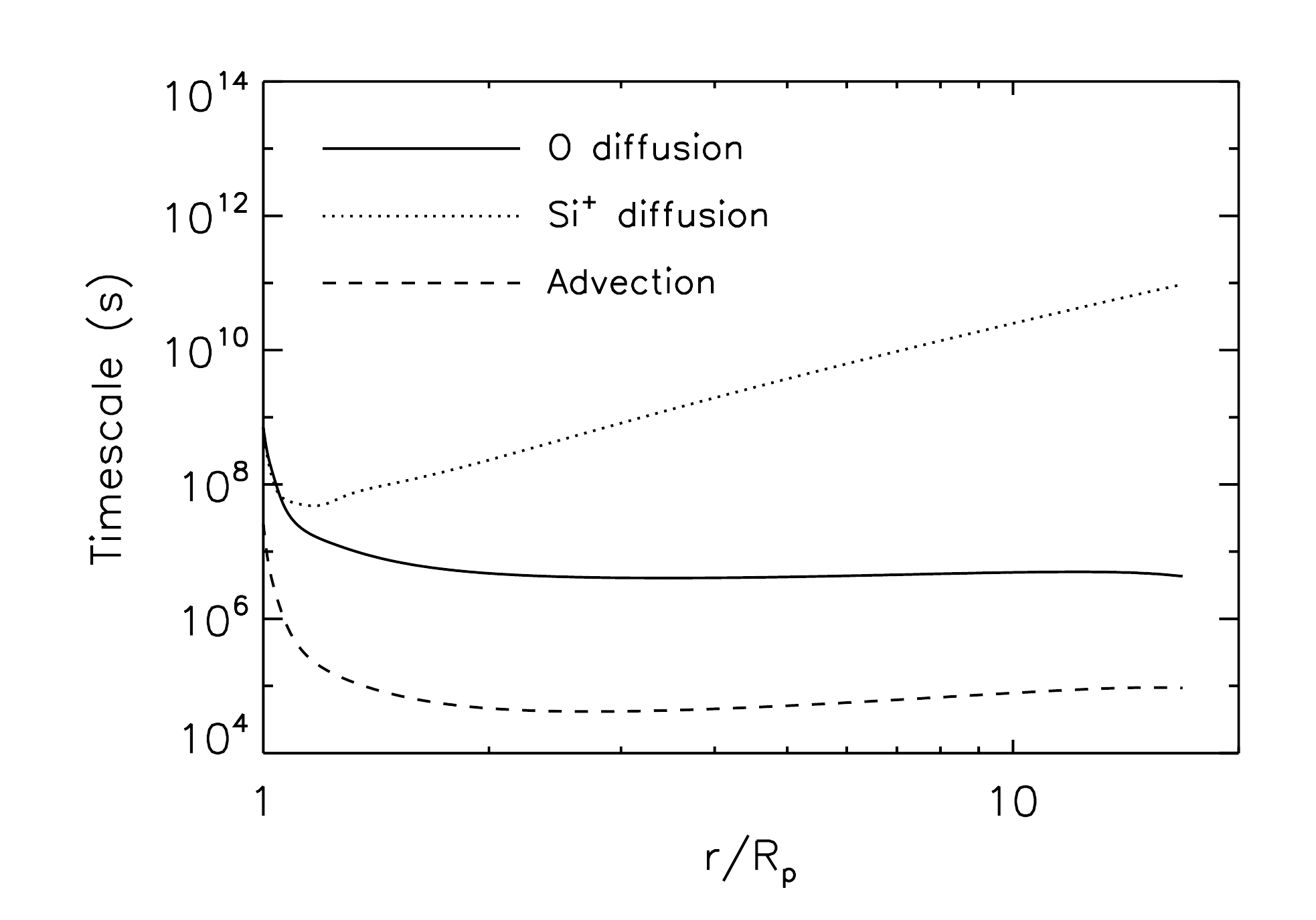}
  \caption{Timescales for diffusion $\tau_D = H^2/D_{s}$ of O and Si$^+$, and for advection $\tau_v = H/v$ based on the C2 model.  We calculated the diffusion coefficients in a mixture of H and H$^+$.  The large scale height of the atmosphere and relatively high collision frequencies mean that diffusion is not significant ($\tau_D/\tau_v = v H/D_s >> 1$) in the thermosphere of HD209458b.}
  \label{fig:difftimes}
\end{figure}

Figure~\ref{fig:difftimes} compares the timescale for diffusion $\tau_D$ for O and Si$^+$ with the timescale for advection $\tau_v$ based on the C2 model.  In both cases, $\tau_D >> \tau_v$ and thus diffusion is not significant.  This implies that there are no significant velocity differences between heavy atoms and hydrogen.  We note that Coulomb collisions of doubly ionized species with H$^+$ are more frequent than collisions between a singly ionized species and H$^+$.  Thus diffusion is even less significant for a species like Si$^{2+}$.  We verified these results from our simulations by switching diffusion off in the model and rerunning the C2 model.  As a result the density of the heavy atoms and ions increased slightly at high altitudes, but the differences are not significant -- the results were nearly identical to the density profiles of the original simulation.  

We note that the atmosphere can also be mixed by vertical motions associated with circulation that are sometimes parameterized in one-dimensional models by the eddy diffusion coefficient $K_{zz}$ \citep[e.g.,][]{moses11}.  This mechanism is efficient in bringing the heavy elements to the lower thermosphere but it is unlikely to mix the atmosphere up to 3 $R_p$ and beyond.  Also, there is no generally accepted method of estimating the degree of global mixing based on circulation models, and most circulation models for EGPs do not adequately treat the relevant energy deposition and forcing mechanisms in the upper atmosphere.  Thus there is considerable uncertainty over the values of $K_{zz}$ and rapid escape is a much more likely explanation for the lack of diffusive separation on HD209458b.  In fact, the calculations of \citet{koskinen07} show that the temperature in the thermosphere of planets such as HD209458b is high enough to practically guarantee an effective escape rate.  The only way to prevent this is to provide enough radiative cooling to offset most of the XUV flux, but there are no radiative cooling mechanisms efficient enough to achieve this in a thermosphere composed of atoms and ions.       

As we noted above, the temperatures of the electrons, ions, and neutrals are roughly equal in the thermosphere of HD209458b.  In order to show this, we assumed that photoelectrons share their energy with thermal electrons, which then share this energy further with ions and neutrals.  We also assumed that the collisions frequencies between the species are higher than the timescale for advection.  If the velocity differences between the species are negligible, the steady state 5-moment energy equations for thermal electrons and ions \citep{schunk00} can be used to obtain the following approximations\footnote{Note that conduction and viscosity are not important in the thermosphere of HD209458b.}:
\begin{eqnarray}
T_e - T_i&\approx&\frac{1}{3} \frac{m_i}{m_e} \frac{q_R}{k n_e \nu_{ei}} \label{eqn:temp1} \\
T_i - T_n&\approx&\frac{1}{3} \frac{m_i + m_n}{m_i} \frac{q_R}{k n_i \nu_{in}} \label{eqn:temp2}
\end{eqnarray} 
where $q_R$ is the volume heating rate, and $\nu_{ei}$ and $\nu_{in}$ are the electron-ion and ion-neutral momentum transfer collision frequencies, respectively.

We calculated the temperature differences for H, H$^+$, and electrons based on the C2 model.  The difference between the electron and ion temperatures decreases with altitude and is mostly less than 2 K.  The difference between the ion and neutral temperatures, on the other hand, increases with altitude.  The ion temperature is approximately 10 K higher than the neutral temperature near 5 $R_p$ and the difference reaches 150 K at 16 $R_p$.  In both cases, the temperature differences are negligible compared to the temperature of the thermosphere.  Further, the timescale for advection in the C2 model is always significantly longer than the relevant collision timescales.  Thus we have shown that $T_e \approx T_i \approx T_n$ and that equations (\ref{eqn:temp1}) and (\ref{eqn:temp2}) are approximately valid.   

\section{Discussion and Conclusions}
\label{sc:discussion} 

We have constructed a new model for the upper atmosphere of HD209458b in order to explain the detections of H, O, C$^+$, and Si$^{2+}$ at high altitudes around the planet \citep{vidalmadjar03,vidalmadjar04,linsky10}.  There are many different interpretations of the observed transits in the H Lyman~$\alpha$ line \citep{vidalmadjar03,benjaffel07,benjaffel08,holstrom08,koskinen10}, and these interpretations rely on results from models of the upper atmosphere that are based on many uncertain assumptions \citep[see Section~\ref{subsc:gendep} and][for a review]{koskinen10}.  Also, the detection of heavy atoms and ions in the thermosphere is not without controversy, and the detection of Si$^{2+}$ is particularly intriguing.  Thus these observations present several interesting challenges to modelers.  

The observed transit depths are large, and substantial abundances of the relevant species are required to explain the observations.  However, on every planet in the solar system heavier species are removed from the thermosphere by molecular diffusion and doubly ionized species are not commonly observed.  Also, the observations imply that H and O remain mostly neutral in the thermosphere while C and Si are mostly ionized at a relatively low altitude.  Hydrodynamic models coupled with chemistry and thermal structure calculations are required to explain the detection of these species in the upper atmosphere and the differences between their density profiles.  Ours is the first such model that benefits from repeated detections of both neutral atoms and ions to constrain the composition and temperature.

\citet{koskinen10} demonstrated that the H Lyman~$\alpha$ transit observations \citep{benjaffel07,benjaffel08} can be explained with absorption by H in the thermosphere if the base of the hot layer of H is near 1 $\mu$bar, the mean temperature within the layer is about 8,250 K, and the atmosphere is mostly ionized above 3 $R_p$.  These parameters are based on fitting the data with a simple empirical model of the upper atmosphere.  The density and temperature profiles from our new hydrodynamic model agree qualitatively with these constraints, demonstrating that the basic assumptions of \citet{koskinen10} are reasonable.  This confirms once again that a comet-like tail \citep{vidalmadjar03} or energetic neutral atoms \citep{holstrom08} are not necessarily required to explain the H~Lyman~$\alpha$ observations.  

In line with recent results by \citet{moses11} and the empirical constraints mentioned above, we used a photochemical model of the lower atmosphere to show that H$_2$ dissociates near the 1 $\mu$bar level.  Above this level, the lack of efficient radiative cooling and strong stellar EUV heating lead to high temperatures.  We constrained the range of possible mean (pressure averaged) temperatures based on the average solar flux by using the hydrodynamic model to calculate temperatures with different heating efficiencies.  For net heating efficiencies between 0.1 and 1, the mean temperature below 3 $R_p$ varies from 6,000 K to 8,000 K.  This means that 8,000 K is a relatively strict upper limit on the mean temperature if the XUV flux of HD209458 is similar to the corresponding flux of the sun.

A mean temperature of 8,250 K estimated from the observations implies the presence of an additional non-radiative heat source, or that the XUV flux from HD209458 is higher than the average solar flux.  Given that our best estimate of the net heating efficiency is 0.44 (see Section~\ref{subsc:cecchieff}), the XUV flux of H209458 would have to be 5--10 times higher than the average solar flux to cause a mean temperature of 8,250 K (see Section~\ref{subsc:gendep}).  If the mean XUV flux of HD209458 is generally higher than the solar flux and the observations took place during stellar maximum, such an enhancement is not impossible.  This would also lead to higher outflow velocity and mass loss rate.  However, the uncertainty in the observed transit depths is also large \citep{benjaffel08,benjaffel10}, and it can accommodate a range of temperatures.  Therefore our reference model C2 with a mean temperature of 7,200 K also agrees qualitatively with the empirical constraints.  In this respect, it is interesting to note that with 100x solar flux, the mean temperature is still only 9,800 K.  Temperatures significantly higher than 8,000 K \citep[e.g.,][]{benjaffel10} therefore imply a strong non-radiative heat source.            

In contrast to the mean temperature, the velocity and details of the temperature profile depend strongly on the heating efficiency and stellar flux (see Section~\ref{subsc:gendep}).  They are also sensitive to the upper and lower boundary conditions.  This explains the large range of temperature and density profiles predicted by earlier models that arise from different boundary conditions and assumptions about the stellar flux, radiative transfer, and heating efficiencies.  The differences highlight the need for accurate thermal structure calculations that are constrained by the available observations.  These calculations are important because the density profiles of the detected species depend on the temperature and velocity profiles, and inappropriate assumptions made by the models can bias the interpretation of the observations.

In the absence of stellar gravity, the location of the sonic point and the outflow speed also depend on the heating efficiency.  As the heating efficiency increases from 0.1 (in models with the average solar flux), the high altitude temperature increases and the sonic point moves to lower altitudes, reaching down to 4 $R_p$ with a net heating efficiency of 1.  We found that supersonic solutions are possible as long as there is significant heating over a large altitude range above the temperature peak.  This conclusion is supported both by the hydrodynamic model and new constraints from kinetic theory \citep{volkov11a,volkov11b}.  However, the isentropic sonic point of the C2 model is located above the model domain.  In principle, this is an interesting result but it should be treated with caution.  We used parameterized heating efficiencies for low energy photons, and the location of the sonic point is very sensitive to the temperature profile.  Also, the stellar tide can enhance the escape rates at the substellar and antistellar points.  We did not include this effect because it may produce horizontal flows that cannot be modeled in 1D.

As long as the upper boundary is at a sufficiently high altitude, we found that the results based on the outflow boundary conditions and modified Jeans conditions are identical (see Section~\ref{subsc:boundaries}).  This shows that our simulations are roughly consistent with kinetic theory.  An agreement between these two types of boundary conditions on HD209458b is an interesting theoretical result.  It shows that the boundary conditions for hydrodynamic escape are appropriate in this case.  However, an upper boundary at 16 $R_p$ or higher is not necessarily justified for other reasons because we did not consider the effect of the possible planetary magnetic field, interaction of the atmosphere with the stellar wind, or horizontal transport \citep[e.g.,][]{stone09,trammell11}.

We chose an upper boundary at a high altitude in order to preserve the integrity of the solution in our region of interest below 5 $R_p$.  The purpose of this work is to model energy deposition and photochemistry in this region.  These aspects are often simplified in more complex models to a degree that it may be difficult to separate the effect of multiple dimensions and other complications from differences arising simply because of different assumptions about heating efficiencies and chemistry.  Also, the uncertainty in the observations does not necessarily justify the introduction of more free parameters to the problem until the basic properties of the thermosphere are better understood.  However, technically we do not consider our solutions to be accurate far above 3--5 $R_p$.  Instead, our results provide robust lower boundary conditions for more complex multidimensional models that characterize the atmosphere outside the Roche lobe of the planet.  Results from such models can then be used to constrain the upper boundary conditions of the 1D models further. 

In order to model the density profiles of the detected species in the ionosphere, we assumed solar abundances of the heavy elements \citep{lodders03}, although this assumption can be adjusted as required to explain the observations (Paper II).  As we already stated we found that H$_2$, H$_2$O, and CO dissociate above the 1 $\mu$bar level, releasing H, O, and C to the thermosphere \citep[see also][]{moses11}.  We note that the detection of Si$^{2+}$ in the upper atmosphere implies that silicon does not condense into clouds of forsterite and enstatite in the lower atmosphere as argued by e.g., \citet{visscher10}.  The dominant Si species is then SiO, which dissociates at a similar pressure level as the other molecules.  In fact, practically all molecules dissociate below 0.1 $\mu$bar.  This leads to an important simplification in hydrodynamic models of the thermosphere.  The complex chemistry of molecular ions does not need to be included as long as the lower boundary is above the dissociation level. 

We found that the H/H$^+$ transition occurs near 3 $R_p$ or, depending on the velocity profile, at even higher altitudes.  The O/O$^+$ transition is coupled to the H/H$^+$ transition through charge exchange reactions.  Thus both H and O are mostly neutral up to the boundary of the Roche lobe at 3--5 $R_p$.  In contrast, C is ionized near the 1 $\mu$bar level and C$^+$ is the dominant carbon species in the thermosphere.  Si is also ionized near the 1 $\mu$bar level, and the balance between Si$^+$ and Si$^{2+}$ is determined by charge exchange with H$^+$ and H, respectively.  Si$^+$ is the dominant silicon ion below 5 $R_p$ but the abundance of Si$^{2+}$ is also significant.  We found that neutral heavy atoms are dragged to the thermosphere by the escaping H, while heavy ions are transported efficiently by the escaping H$^+$.  Thus the advection timescale is much shorter than the diffusion timescale of the detected species, and diffusive separation does not take place in the thermosphere.  We also verified that the neutral, ion, and electron temperatures are roughly equal.  

Taken together, these results imply that the thermospheres of close-in EGPs can differ fundamentally from the gas giant planets in the solar system.  For instance, the thermosphere of HD209458b is composed mainly of atoms and atomic ions, and diffusive separation of the common heavy species is prevented by the escape of H and H$^+$.  It is important to note, however, that results such as these cannot be freely generalized to other extrasolar planets.  As in the solar system, each planet should be studied separately.  For instance, the dissociation of molecules depends on the temperature profile that is shaped by the composition through radiative cooling and stellar heating.  The mass loss rate and escape velocity, that determine whether diffusive separation takes place or not, depends on the escape mechanism that again depends on the temperature and composition of the upper atmosphere.  The results from different models can only be verified by observations that are therefore required for multiple planets if we are to characterize escape in different systems and under different conditions.\\
    
We are grateful to A. Volkov for reading the manuscript and providing useful feedback.  We thank H. Menager, M. Barthelemy, J.-M. Grie\ss meier, N. Lewis, D. S. Snowden, and C. Cecchi-Pestellini for useful discussions and correspondence.  We also acknowledge the "Modeling atmospheric escape" workshop at the University of Virginia and the International Space Science Institute (ISSI) workshop organized by the team "Characterizing stellar and exoplanetary environments" for interesting discussions and an opportunity to present our work.  The calculations for this paper relied on the High Performance Astrophysics Simulator (HiPAS) at the University of Arizona, and the University College London Legion High Performance Computing Facility, which is part of the DiRAC Facility jointly funded by STFC and the Large Facilities Capital Fund of BIS.  SOLAR2000 Professional Grade V2.28 irradiances are provided by Space Environment Technologies.

\end{document}